\documentclass[final]{IEEEtran}
\usepackage{amsthm,amssymb,graphicx,multirow,amsmath,color,amsfonts}
\usepackage[update,prepend]{epstopdf}
\usepackage[noadjust]{cite}
\usepackage[latin1, utf8]{inputenc}
\usepackage{tikz}
\usepackage{bbm} 
\usepackage{pdfpages}
\usepackage{tabulary}
\usepackage{multirow}
\usepackage{comment}
\usepackage{subfigure}
\usepackage{float}

\usepackage[titlenumbered,ruled]{algorithm2e}
\usepackage[noend]{algpseudocode}

\def\nb0{{\mathbf{0}}}
\def\nb1{{\mathbf{1}}}

\newtheorem{lemma}{Lemma}

\newtheorem{definition}{Definition}

\newtheorem{theorem}{Theorem}

\newtheorem{remark}{Remark}

\def\argmin{\operatorname{arg~min}}

\allowdisplaybreaks 
\begin{document}
\title{Stochastic Geometry-based Analysis of Multi-Purpose UAVs for Package and Data Delivery}
\author{
	Yujie Qin, Mustafa A. Kishk, {\em Member, IEEE}, and Mohamed-Slim Alouini, {\em Fellow, IEEE}
	\thanks{Yujie Qin and Mohamed-Slim Alouini are with Computer, Electrical and Mathematical Sciences and Engineering (CEMSE) Division, King Abdullah University of Science and Technology (KAUST), Thuwal, 23955-6900, Saudi Arabia
		Arabia. Mustafa Kishk is with the Department of Electronic Engineering, National University of Ireland, Maynooth, W23 F2H6, Ireland. (e-mail: yujie.qin@kaust.edu.sa; mustafa.kishk@mu.ie; slim.alouini@kaust.edu.sa).} 
	
}
\date{\today}
\maketitle
\date{\today}
\maketitle
\begin{abstract}
	Using drones for communications and transportation is drawing great attention in many practical scenarios, such as package delivery and providing additional wireless coverage. However, the increasing demand for UAVs from industry and academia will cause aerial traffic conflicts in the future. This, in turn, motivates the idea of this paper: multi-purpose UAVs, acting as aerial wireless data relays and means of aerial transportation simultaneously, to deliver packages and data at the same time. This paper aims to analyze the feasibility of using drones to collect and deliver data from the Internet of Things (IoT) devices to terrestrial base stations (TBSs) while delivering packages from warehouses to residential areas. We propose an algorithm to optimize the trajectory of UAVs to maximize the size of collected/delivered data while minimizing the total round trip time subject to the limited onboard battery of UAVs. Specifically, we use tools from stochastic geometry to model the locations of the IoT clusters and the TBSs and study the system performance with respect to energy efficiency, average size of collected/delivered data, and package delivery time. Our numerical results reveal that multi-functional UAVs have great potential to enhance the efficiency of both communication and transportation networks.
\end{abstract}
\section{Introduction}
With technological progress, unmanned aerial vehicles (UAVs, also known as drones) have recently drawn increasing interest from both industry and academia and are expected to play an essential role in potentially enhancing the performance of the next generation wireless networks \cite{sekander2018multi,mozaffari2019tutorial,8579209,zeng2016wireless,khosravi2021multi}. Unlike the traditional terrestrial base stations (TBSs), UAVs are more flexible and can easily satisfy the dynamic traffic demands by optimizing their locations in real-time. Since they can adjust their altitudes, they are more likely to establish line-of-sight (LoS) links with ground users \cite{al2014optimal,8833522}. At places where the users exhibit a certain degree of spatial clustering, and the density of active users varies with time, UAVs can rapidly function as aerial base stations (ABSs) and offer an additional dynamic capacity.

In addition, for the Internet of Things (IoT) devices specifically, UAV-evolved networks are more suitable. On one hand, the transmit power of IoT devices is much lower compared to mobile users owing to their limited energy. Hence, they require efficient data transfer links. On the other hand, IoT devices do not require data transmission all the time. They can collect a large amount of data and transfer them together, say once a day. In this case, UAVs are considered as a competitive candidate to serve these devices \cite{motlagh2016low}, since they can first wirelessly charge IoT devices \cite{gupta2021pricing}, establish LoS links, collect data efficiently based on the dynamic demand of the IoT devices and hence, prolong the lifetime of the network.

Besides communication advantages, another important application of UAVs is on-demand transport of packages. Delivery applications include package deliveries, such as last-mile deliveries, remote areas, and door-to-door express deliveries. Drones are also expected to deliver medical products, such as drone-based contact-less COVID-19 diagnosis and testing \cite{naren2021iomt}, vaccines, and blood, which significantly improve the quality of medical services. Drone-based flying taxis are being tested and expected to begin commercial operations soon. Compared with a traditional truck delivery system, a drone-based system may be faster due to the lower traffic and ability to avoid low altitude obstacles and have a lower cost per mile to operate \cite{otto2018optimization}.

Generally, drones are typically designed to be dedicated to a single purpose. However, several demands of personal and commercial applications will cause heavy traffic conflicts in future aerial networks \cite{khosravi2021multi} and it is not energy or space-efficient for all the operators to have their own physical infrastructure or own dedicated UAVs. One of the most straightforward solutions is sharing: drones can be designed more flexibly to finish multiple tasks at the same time, such as delivering a package while providing coverage \cite{khosravi2021multi} or transferring data. This paper aims to investigate the possibilities of delivering packages and data simultaneously. With that being said, we consider an energy-limited rotary-wing UAV to finish multiple tasks simultaneously. Specifically, we use stochastic geometry and optimization tools to analyze the system performance, such as energy consumption, delivery time, total round trip time, and maximum collected/delivered data, in a general case.
\subsection{Related Work}
Literature related to this work can be categorized into: (i) UAV-enabled package delivery, (ii) UAV-enabled communication networks, and (iii) stochastic geometry-based analysis of UAV networks. A brief discussion on related works is provided in the following lines.

{\em UAV-enabled package delivery.} Considering drone for last-mile delivery is a recent hot topic. A creation of technology road mapping for drones, used by Amazon for their latest service Amazon Prime Air, was provided in \cite{singireddy2018technology,welch2015cost}. Authors in \cite{yoo2018drone} examined the determinants of the customer adoption of drone delivery. Authors in \cite{agatz2018optimization} studied the traveling salesman problem. Compared to truck-only delivery, their experiment showed that substantial savings are possible by using drone-based delivery. Besides packages, drones can also be used in delivering medical supplies \cite{ling2019aerial}. In \cite{8701196}, authors mentioned that Zipline delivers blood as well as other urgently medical supplies to hospitals and clinics every day in Rwanda. Under the current situation of COVID-19, drones are beneficial in handling the pandemic since they can cover a large area and provide data such as thermal image and patient identification \cite{kumar2021drone} and contactlessly deliver the test kit to the patient having a high likelihood of infected \cite{naren2021iomt}. Moreover, drone-based package delivery are expected to be cost-competitive, fast and conveniently accessible compared to traditional ground-based delivery in or near urban regions \cite{otto2018optimization,yoo2018drone,agatz2018optimization,rabta2018drone,poikonen2017vehicle,nedjati2016post}.
In addition to package delivery, areal passenger transportation technology has also been progressing lately (flying taxi) \cite{zhu2019pre}.

{\em UAV-enabled communication networks.} Authors in \cite{otto2018optimization} comprehensively surveyed the applications of UAVs and related approaches to optimize their operations, such as path planning of area coverage, routing for a set of locations, planning of data gathering, and recharging in wireless sensor networks. Authors in \cite{gupta2021pricing} envisaged the idea of UAVs dedicated for wireless charging of sensor networks. They presented a suitable pricing model based on a game-theoretic approach. Similarly, IoT devices are also energy-limited. Authors in \cite{8842600} maximized the number of served IoT devices by jointly optimizing the trajectory of a UAV and radio resource allocation. Besides optimizing the trajectory of a single UAV, network architecture and optimal number of UAVs were analyzed, and a low latency routing algorithm was designed based on the network architecture and partial location information in \cite{8599015}. Reliable uplink communications between UAVs and IoT devices with minimum transmit power were investigated in \cite{8038869}. Using the proposed approach, reliability can be highly enhanced while the transmit power is sharply reduced.
Authors in \cite{8489918} investigated a throughput maximization problem in a UAV-enabled network, where UAV first wireless charges users in downlink then the user uses the harvested energy to transmit in uplink. In \cite{7486987}, the authors concluded that UAVs' altitude and the beamwidth of the antenna should be adjusted properly to mitigate interference. The possible application of UAV-enabled Intelligent transportation system (ITS) was analyzed in \cite{7876852}, where authors studied the applications of UAV-enabled ITS, UAV deployment, and discussed the security challenges. UAVs' transmission completion time was firstly analyzed in \cite{8902102} and the authors developed an algorithm to efficiently minimize the time. A one/two-UAV deployment strategy was analyzed in \cite{chargingpad}, where the authors computed the energy efficiency and showed that an optimal cluster pair density exists. Authors in \cite{zhou2019secure} studied the UAV-enabled mobile edge computing systems and showed the tradeoff between security and latency of such a system.
Authors in \cite{8998329} proposed a novel system model for cooperative transmission for UAV users. They considered UAVs as users and having frequent movements, and TBSs serving multiple UAV users.

{\em Stochastic geometry-based analysis of UAV networks.} Stochastic geometry is a strong mathematical tool that enables characterizing the statistics of various large-scale wireless networks~\cite{7733098,6524460}. UAV-enabled fair shared spectrum access was analyzed in the coexistence of cellular users and IoT devices in \cite{9103583} to maximize their energy efficiency. Authors in~\cite{8833522,galkin2019stochastic,alzenad20173} studied a heterogeneous network composed of TBSs and UAVs, in which the locations were modeled by two independent PPPs. Downlink coverage probability and average data rate for the considered setup were derived after accurately characterizing the Laplace transform of the interference coming from both aerial and terrestrial BSs. Another commonly used point process, 'Matern cluster process (MCP)', was used in~\cite{7809177,9773146,qin2021influence,qin2020performance} to analyze the system performance, where users were assumed to be uniformly located within the user cluster.
A stochastic geometry-based moving aerial base stations  (ABSs) network was designed in \cite{8681266}. The authors studied two special cases and shown that moving ABSs provide both mobility and coverage benefits.
In \cite{9095281}, the authors proposed a novel system where aerial users periodically transmit with BSs while moving via a given trajectory and observed that the minimum height of exceeding target rates depends on the distance to the BSs.
Authors in \cite{9013645} proposed a stochastic geomery-based mobility model for  UAV-enabled cellular networks, considered two mobility models, and compared their models with some more complicated model, where UAVs move in nonlinear trajectories.

While the existing literature mainly focus on single application of UAV-enabled network, there is few work about integrating these functions together \cite{khosravi2021multi}. While authors in \cite{khosravi2021multi} optimized the trajectory of UAVs when delivering packages to provide uniform coverage for the whole area, our UAVs acting as aerial wireless data relays and means of aerial transportation to deliver packages and collect/deliver data from IoT clusters simultaneously.

\subsection{Contribution}
This paper systematically investigates the feasibility of integrating communication and transport functions on a single UAV and its performance. Needless to say, while we choose package and IoT devices data delivery as our main focus, this idea holds for all kinds of applications: providing coverage to user clusters, communicating with roadside units or vehicles while delivering packages or even passengers, etc. Besides, we consider all system components to be randomly located to be more realistic and capture the average system performance. More detailed discussions are provided next.

{\em Novel Framework and Performance Metrics.} To solve the problem of heavy traffic conflicts in future aerial networks, we propose a novel system where UAVs are multi-functional: transferring data from IoT devices to TBSs while delivering packages. Given these two goals of this UAV network, we systematically investigate the network performance from each perspective of network components by defining delivery efficiency and computing maximum collected/delivered data and the minimum round trip time subject to UAVs' limited energy.

{\em UAVs' optimal trajectory.} 
From the perspective of operators, we optimize the trajectory of UAVs to maximize the collected/delivered
IoT data and minimize the time spent in the whole trip (e.g., finish both assignments: deliver the package and transfer the data) within the limited battery energy. Unlike existing literature, we use tools from stochastic geometry and model the locations of IoT clusters and TBSs by two independent PPPs, which is more realistic and has no restrictions on the locations. That is, our proposed algorithm of optimizing the trajectory fits different kinds of scenarios.

{\em System-Level Insights.} Since we consider all the nodes to be randomly located, our results are based on average system performance. Our results reveal that it is more time and energy-efficient for UAVs to finish multi-tasks simultaneously. We show that the idea of integrating the aforementioned functions on single UAVs works well in general cases, which opens interesting topics in future UAV networks.

\section{System Model}
\begin{figure}
	\centering
	\includegraphics[width= 1\columnwidth]{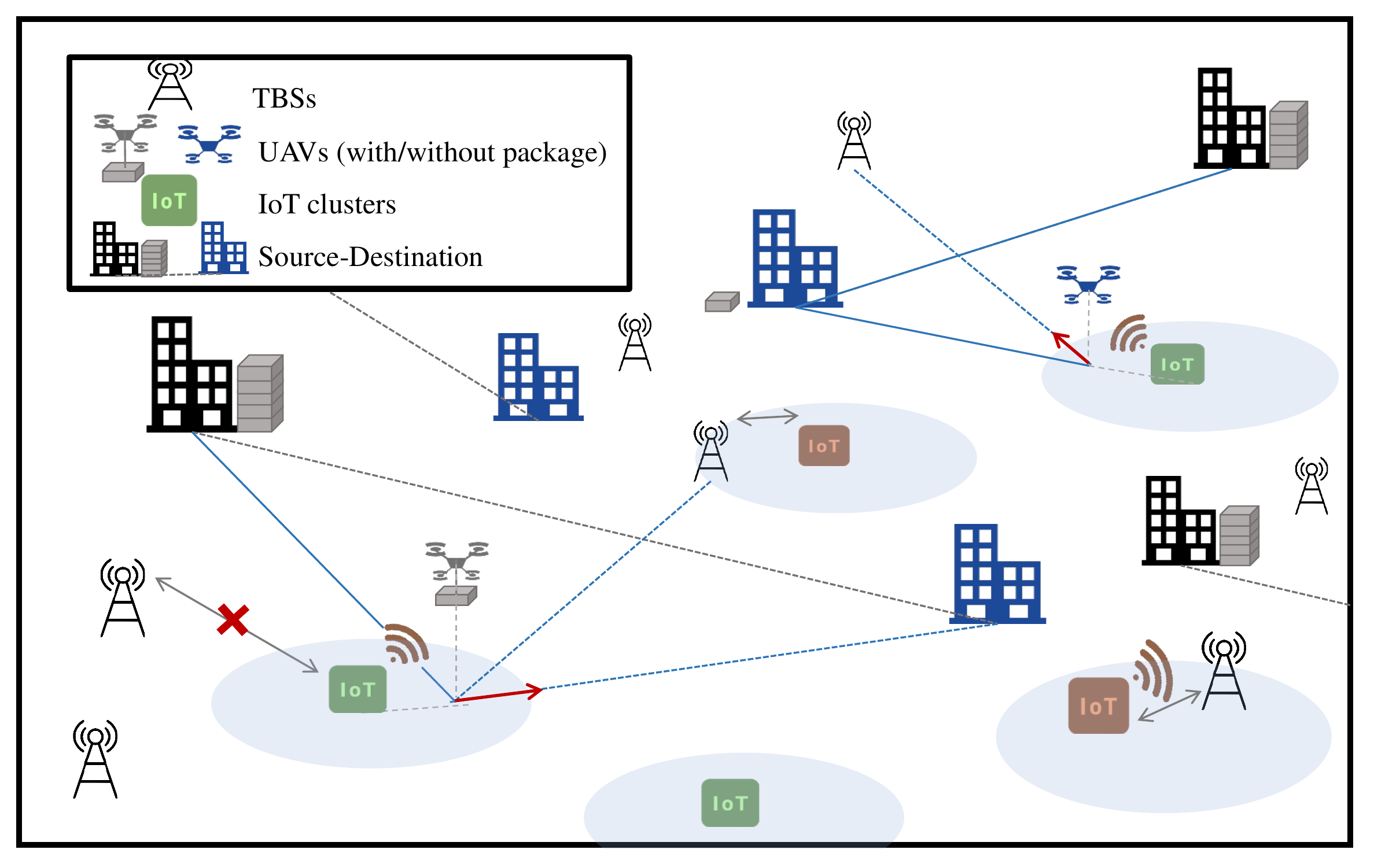}
	\caption{Illustration of system model.}
	\label{fig_sys1}
\end{figure}

Given that IoT devices require reliable communication channels, are energy limited, and usually, TBSs are too far and consume a large amount of energy,
we consider a UAV delivery system that can simultaneously help IoT devices to communicate with TBSs, as shown in Fig. \ref{fig_sys1}. With that being said, while UAVs deliver packages from the warehouses to residential areas (sources to the destinations), we consider them to be able to collect data from IoT devices and then forward it to the nearest TBS along the route, and the distance between the nearest TBS to the UAV trajectory is $R_b$ (more details about the distance are provided in Section \ref{setion_Distance}).
The locations of source and destination (S-D) pairs are modeled by Poisson bipolar networks, in which each source has a destination at a distance $L_2$ in a random orientation. The UAV starts at $S$ with a package and drops off the package when it arrives at $D$.
The locations of TBSs and IoT cluster centers are modeled by two independent Poisson point processes (PPPs) $\Phi_t$ and $\Phi_i$, with different densities $\lambda_t$ and $\lambda_i$, respectively. We consider that UAVs only provide service (data collection and delivery) to the IoT clusters which are far from TBSs and their data rate (when communicating with TBSs directly without UAVs) is lower than a predefined threshold $c_t$.

\subsection{Power Consumption}
Consider the UAV's onboard battery limitation to be the main system's bottleneck. UAVs rely on their internal battery for power supply, hence, the amount of time they can fly/hover, providing service as well as the weight of package they can carry is limited. Generally, the total power consumption of UAVs is composed of communication-related power and propulsion-related power. The power consumption model of this work is based on \cite{zeng2019energy}. 

The power consumption of a rotary-wing UAV, which is sensitive to the payload and size, is given by
\begin{align}
	\begin{aligned}
		p(V)=& P_{0}\left(1+\frac{3 V^{2}}{U_{\text {tip }}^{2}}\right)+P_{i}\left(\sqrt{1+\frac{V^{4}}{4 v_{0}^{4}}}-\frac{V^{2}}{2 v_{0}^{2}}\right)^{1 / 2}\nonumber\\
		&+\frac{1}{2} d_{0} \rho s A V^{3},
	\end{aligned}
\end{align}
where,
\begin{align}
	P_{0} &= \frac{\delta}{8} \rho s A \Omega^{3} R^{3},\nonumber\\
	P_{i} &= (1+k) \frac{W^{3 / 2}}{\sqrt{2 \rho A}},
\end{align}
in which $W$ is the total weight of UAVs (the sum of both aircraft and the average weight of the package $\bar{w}$), $V$ is the velocity of UAVs, $\rho$ is air density, $R$ is rotor radius, $A$ is the area of rotor disc, $v_0$ is mean rotor induced velocity, $U_{\rm tip}$ denotes the tip speed, $s$ is rotor solidity, $\Omega$ is blade angular velocity, $R$ is rotor radius, $k$ is incremental correction factor and $\delta$ is profile drag coefficient.

Let $p_{m}(V)$ and $p_{s}(V)$ be the  motion- and service-related power of UAVs without carrying packages, and $p_{mp}(V)$ and $p_{sp}(V)$ be the  motion- and service-related power of UAVs with packages. We consider that UAVs use the optimal velocities to minimize the power consumption when traveling, and let $v$ and $v_p$ be the optimal velocities without/with package, respectively. Besides, the service-related power of UAVs is composed of hovering and communication power, and the velocity of UAVs is 0 when hovering. Hence, in the following text, we simplify the notations and use $p_{\{m,mp,s,sp\}}$ since all the velocities are constant.

\subsection{Communication Channel and Time Consumption}

The communication channels between  UAVs and (i) IoT devices (I2U), and  (ii) TBSs (U2B) are modeled as Nakagami-m fading channels. 
Given the horizontal distances between the serving UAV and IoT device, TBS are $R_{\rm i2u}$ and $R_{\rm u2b}$, respectively, the received power of UAVs from IoT devices is given by
\begin{align}
	&p_{\mathrm{i}}(R_{\rm i2u})=\nonumber\\
	&\left\{\begin{array}{l}
		p_{\mathrm{i,l}}(R_{\rm i2u})=\eta_{\mathrm{l}} \rho_{\mathrm{i}} G_{\rm l} D_{\rm i2u}^{-\alpha_{\mathrm{l}}}, \text { in the case of } \mathrm{LoS} \\
		p_{\mathrm{i,n}}(R_{\rm i2u})=\eta_{\mathrm{n}} \rho_{\mathrm{i}} G_{\mathrm{n}} D_{\rm i2u}^{-\alpha_{\mathrm{n}}}, \text { in the case of } \mathrm{NLoS}
	\end{array}\right.
\end{align}
where $D_{\rm i2u} = \sqrt{h^2+R_{\rm i2u}}$ and $h$ is the UAVs' altitude. Similarly, the received power of TBSs from UAVs is given by
\begin{align}
	&p_{\mathrm{u}}(R_{\rm u2b})=\nonumber\\
	&\left\{\begin{array}{l}
		p_{\mathrm{u,l}}(R_{\rm u2b})=\eta_{\mathrm{l}} \rho_{\mathrm{u}} G_{\rm l} D_{\rm u2b}^{-\alpha_{\mathrm{l}}}, \text { in the case of } \mathrm{LoS} \\
		p_{\mathrm{u,n}}(R_{\rm u2b})=\eta_{\mathrm{n}} \rho_{\mathrm{u}} G_{\mathrm{n}} D_{\rm u2b}^{-\alpha_{\mathrm{n}}}, \text { in the case of } \mathrm{NLoS}
	\end{array}\right.
\end{align}
where $D_{\rm u2b} = \sqrt{h^2+R_{\rm u2b}}$, $\eta_{\mathrm{l}}$ and $\eta_{\mathrm{n}}$ are the mean additional losses for LoS and NLoS transmissions, $\alpha_{\rm l}$ and $\alpha_{\rm n}$ are the path loss of LoS and NLoS links, respectively, $G_{\rm l}$ and $G_{\rm n}$ denote the fading gains that follow Gamma distribution with shape and scale parameters $(m_{\rm l},\frac{1}{m_{\rm l}})$ and $(m_{\rm n},\frac{1}{m_{\rm n}})$, $\rho_{\mathrm{i}}$ and $\rho_{\mathrm{u}}$ are the transmit power of IoT devices and UAVs, respectively. The occurrence probability of LoS links and NLoS links between UAVs and serving targets (IoT devices or TBSs) are functions of Euclidean distance $r$, which are given in \cite{al2014optimal} as
\begin{align}
	P_l(r) &= \frac{1}{1+a\exp(-b(\frac{180}{\pi}\arctan(\frac{h}{r})-a))},\nonumber\\
	P_n(r) &= 1-P_l(r),\label{eq_LoSprob}
\end{align}
where $a$ and $b$ are two environment variables.

The fading channels between IoT devices and TBSs (I2B) are modeled as a special case of Nakagami-m fading, denoted by $G_{\rm g}$, where the shape and scale parameters $(m_{\rm g},\frac{1}{m_{\rm g}})$ are fixed as $(1,1)$. Hence, the received power of TBSs from IoT devices directly is
\begin{align}
	p_{\rm b}(R_{\rm i2b}) = \rho_{\rm i}G_{\rm g}R_{\rm i2b}^{-\alpha_{\rm t}},
\end{align}
where $\alpha_{\rm b}$ is the path loss and $R_{\rm i2b}$ is the related distance.

Consequently, for each of the IoT device and UAV, the maximum achievable rate in bps/Hz is given by
\begin{align}
	C(R) &= \log_2\bigg(1+\frac{p(R)}{\sigma^2}\bigg),
\end{align}
where $C(\cdot)\in\{C_{\rm i2u}(R_{\rm i2u}),C_{\rm u2b}(R_{\rm u2b}),C_{\rm i2b}(R_{\rm i2b})\}$ corresponding to $p(\cdot)\in\{p_{\rm i}(R_{\rm i2u}),p_{\rm u}(R_{\rm u2b}),p_{\rm b}(R_{\rm i2b})\}$, respectively, and $\sigma^2$ is the noise power. 

Let $R_{\rm c2u},R_{\rm u2b}$ and $R_{\rm c2b}$ be the horizontal distances between the IoT cluster center to UAVs, UAVs to TBSs and  IoT cluster center to TBSs, respectively, and let $\bar{C}(\cdot)$ be the average maximum achievable rate (average over all the IoT devices within the cluster). $\bar{C}(\cdot)$ is given by $\bar{C}(R^{'}) = \log_2\bigg(1+\mathbb{E}_{R}[\frac{p(R)}{\sigma^2}]\bigg)$, where $\bar{C}(\cdot)\in\{\bar{C}_{\rm c2u}(R_{\rm c2u}),\bar{C}_{\rm u2b}(R_{\rm u2b}),\bar{C}_{\rm c2b}(R_{\rm c2b})\}$.

 \begin{remark}
 	We provide an explanation for the notations of $R$ and $R^{'}$, $\bar{C}(\cdot)$ and $C(\cdot)$ here. $R$ contains the distances between each of the IoT device to the serving BS, which can be UAV or TBS, $R_{\rm i2u}$ and $R_{i2b}$. $R^{'}$ contains the distances between IoT cluster center to the serving BS, $R_{c2u}$ and $R_{\rm c2b}$. The reason for this notation is, each device has its own data rate and transmission time. However, when we analyze the performance, say time consumption, it is the performance of a cluster. $C(\cdot)$ is the data rate for a device while $\bar{C}(\cdot)$ is the average data rate for an IoT cluster. If the total transmission time of a whole cluster is too long, this cluster needs a UAV to help to collect/deliver data. Besides, the distance $R$ is conditioned on $R^{'}$ (e.g., $R_{\rm i2u}$ is function of $R_{c2u}$, the relations are given in Lemma \ref{lemma_RRprime}). Hence, after we take the expectation of $C(\cdot)$ over $R$,  we obtain $\bar{C}(\cdot)$ which is conditioned on $R^{'}$.
 \end{remark}

\begin{definition}[Time Consumption] \label{Def_TimeCon}
For a certain IoT cluster and UAV to TBS link, by taking the expectation over the channel fading,	the total transmission time is 
\begin{align}
	T(R^{'}) &= \mathbb{E}_{\rm G}\bigg[\frac{M_t}{\bar{C}(R^{'})}\bigg] \stackrel{(a)}{\approx} \mathbb{E}_{\rm G}\bigg[\frac{M_t}{b_w\log_2(1+{\rm SNR|R^{'}})}\bigg],	
\end{align}
where the subscript $G$ denotes the channel fading, the approximation in step $(a)$ follows from we take the expectation of $R$ inside the logarithm operation, and  ${\rm SNR|R^{'}} \triangleq \mathbb{E}_{R}[\frac{p(R)}{\sigma^2}]=\int_{r}\frac{p(r)}{\sigma^2}f_{R|R^{'}}(r){\rm d}r$ (the conditional PDF is given in Lemma \ref{lemma_RRprime}), which is a random variable because we only take the expectation over the distances and the channel fading are still random, $M_t$ is the collected/delivered data, $b_w$ is the bandwidth, and $T(R^{'})\in\{T_{\rm c2u}(R_{\rm c2u}),T_{\rm u2b}(R_{\rm u2b}),T_{\rm c2b}(R_{\rm c2b})\}$ corresponds to each $\bar{C}(\cdot)$ and $p(\cdot)$ mentioned above. 
\end{definition}

As mentioned, we consider that if the average maximum achievable rate of I2B channels of
an IoT cluster is lower than a predefined threshold $c_t$, this IoT cluster requires a UAV to deliver
the data to TBSs. The density of IoT clusters require UAVs is $\lambda_{i}^{'}$, which will be computed later in the paper.

\subsection{Optimal Trajectory}
The objective of this work is to optimize the UAV trajectory to maximize the collected/delivered data between IoT devices and TBSs: data collected from IoT devices and delivered to TBSs.
 Here, we consider that UAVs transfer all the data collected from IoT devices. That is, UAVs modify their route and hovering time to ensure that forward all the data collected from IoT devices to TBSs.

 To be more realistic, we consider that the required collected/delivered data from the IoT cluster side is limited, say $M$. Within the limited energy of the UAV battery, we optimize the UAV trajectory to maximize the collected/delivered data while minimizing the total round trip time. In other words, for a given location of the IoT cluster and TBS, if the UAV can deliver all the data, we minimize the required time. If the UAV cannot, we maximize the collected/delivered data. 

We need to clarify that UAVs do not necessarily go to IoT cluster centers and are exactly above TBSs to communicate. Instead, they can hover at nearby points to provide service, and our goal is to find the optimal hovering points (say, $\mathbf{H_1}$ to collect data from IoT clusters and $\mathbf{H_2}$ to deliver data to TBSs), which minimizes the energy consumption or the round trip time. To simplify the notation, we use $\mathbf{H_1,H_2}$ to denote the locations  and the same apply to $\mathbf{L_{IoT},L_{TBS}}$, the locations of IoT cluster center and TBS, which are used in the following text.

In this case, given the locations of the IoT cluster center and TBS, the objective functions are given as:
\begin{align}
	(\mathcal{P}_1):	& \max_{\rm \mathbf{H_1,H_2}\in \mathbb{R}^2} \frac{M_{t|\mathbf{L_{IoT},L_{TBS}}}}{b_w}\nonumber\\
	& {\rm s.t.}\quad E_t ({\rm \mathbf{H_1,H_2}}) \leq B_{\rm max},
\end{align}
where $M_{t|\mathbf{L_{IoT},L_{TBS}}}$ is the conditional collected/delivered data size, $B_{\rm max}$ is the maximum battery size and $E_t ({\rm \mathbf{H_1,H_2}})$ is the total energy consumed of given trajectory. 

Let $T$ be the time for UAVs to finish the round trip (deliver the package and collect/deliver the data) and $T|\mathbf{L_{IoT},L_{TBS}}$ is the conditional round trip time,
\begin{align}
	(\mathcal{P}_2):	& \min_{\rm \mathbf{H_1,H_2}\in \mathbb{R}^2} T|\mathbf{L_{IoT},L_{TBS}} \nonumber\\
	& {\rm s.t.}\quad  E_{\rm T|\mathbf{L_{IoT},L_{TBS}}} ( \mathbf{H_1,H_2})\leq B_{\rm max},\nonumber\\
	&\qquad \frac{M_{t|\mathbf{L_{IoT},L_{TBS}}}}{b_w} = \frac{M}{b_w},
\end{align}
where $ E_{\rm T|\mathbf{L_{IoT},L_{TBS}}} ( \mathbf{H_1,H_2})$ is the corresponding energy consumption of the trajectory, and $M$ is the total size of the data needed to be collected from IoT devices.

To simplify the selection of parameters, we consider transferred data over bandwidth as one parameter and U2B and I2U channels use the same bandwidth. While our goal is to maximize the collected/delivered data, it refers to maximize the collected/delivered data per Hz.

\begin{definition}[Optimal Hovering Points] 
	\label{Def_hov12}
	Let $\mathbf{H_{1}^{*}, H_{2}^*}$ be the optimal hovering points for UAV to collect and deliver the data: $\mathbf{H_{1}^{*}, H_{2}^*} $ are the solutions to the objective functions $(\mathcal{P}_1)$ and $(\mathcal{P}_2)$.
\end{definition}

Notice that the above optimization problems solve for conditional cases, conditioned on $\mathbf{L_{IoT},L_{TBS}}$, but we are interested in general performance, which is defined below.
\begin{definition}[General Case] 
	\label{Def_generalcase}
	The general system performance is given below,
 \begin{align}
	&\{T,E_{\rm total},\frac{M_{\rm t}}{b_w}\}= \nonumber\\
	& \mathbb{E}\bigg[\{T_{\rm min|\mathbf{L_{IoT},L_{TBS}}},E_{\rm total|\mathbf{L_{IoT},L_{TBS}}} ,\frac{M_{\rm t|\mathbf{L_{IoT},L_{TBS}}}}{b_w}\}\bigg],
\end{align}
where $E_{\rm total}$ is the energy consumption of the optimal path and above expectation is over the locations of IoT clusters and TBSs.
\end{definition}

As our focus in this paper are on package  and data delivery. Here, we define a notation $\xi$ referring to package deliver efficiency to characterize the additional time consumed to collect and deliver data from IoT devices.
\begin{definition}[Delivery Efficiency] Delivery efficiency is defined as a time fraction
	\begin{align}
		\xi = \frac{T_{\rm data}}{T_{\rm nodata}},
	\end{align}
	where $T_{\rm nodata}$ and $T_{\rm data}$ are the delivery time of UAV with/without data collection.
\end{definition}
If $\xi$ is close to $1$, it means that data collection has less impact on the package delivery, thus, high efficiency.

\section{Distance Distribution}\label{setion_Distance}
\begin{figure}
	\centering
	\includegraphics[width= 1\columnwidth]{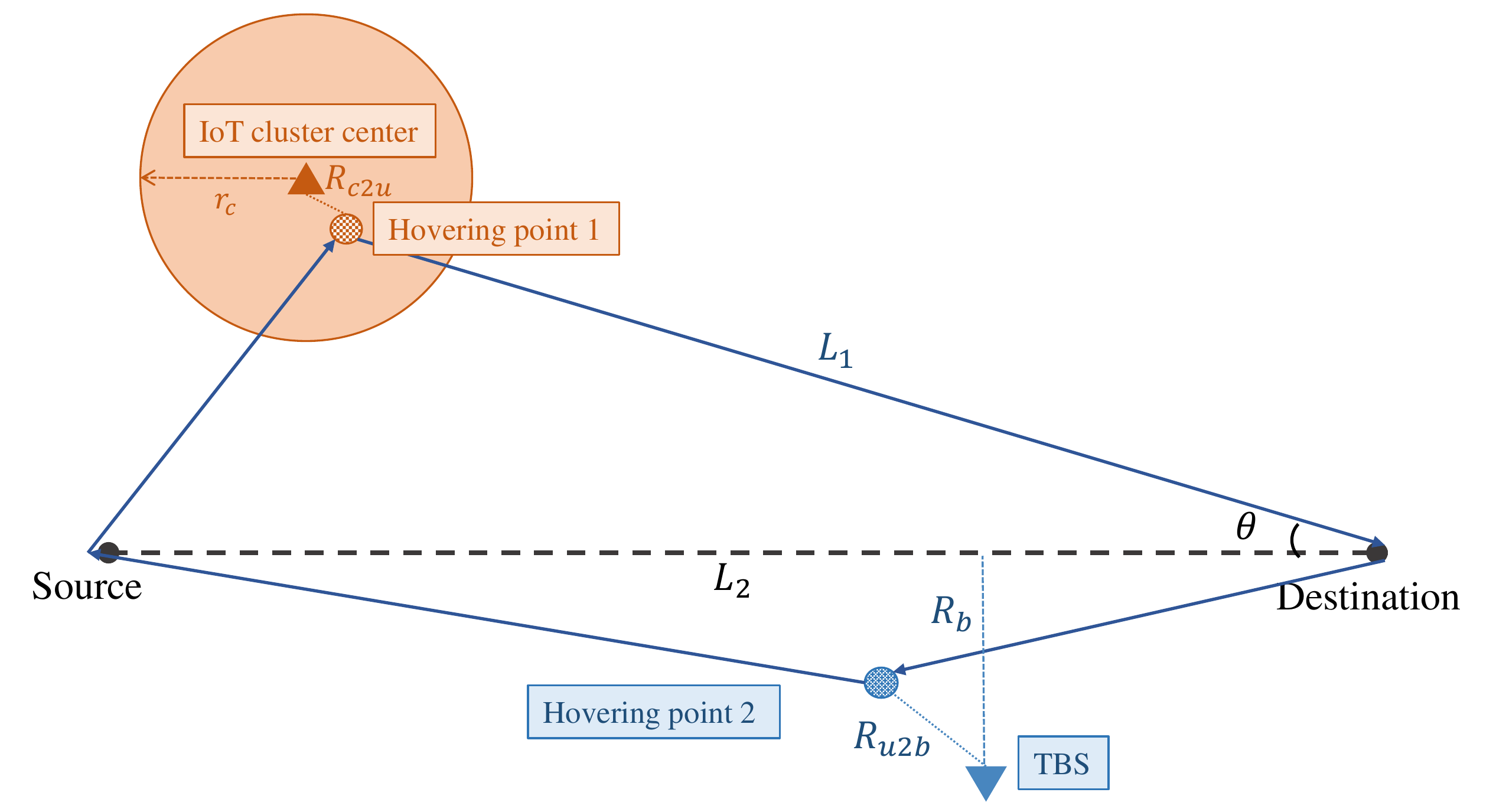}
	\caption{Illustration of the distances.}
	\label{fig_sys2}
\end{figure}

Recall that $R_{\rm c2u}$ and $R_{\rm u2b}$ are the distances between the IoT cluster center to the serving UAVs and the nearest TBS, and $R_b$ is the distance between the nearest TBS to the UAV route, as shown in Fig. \ref{fig_sys2}. Before optimizing the UAV trajectory, the distance distributions are required. Remember that the communicating distance between UAVs and TBSs $R_{\rm u2b}$ is not a random variable since we optimize the UAV trajectory and once we obtain the $\mathbf{H_2^*}$ in Definition \ref{Def_hov12}, $R_{\rm u2b}$ is determined and given by $R_{\rm u2b} = ||\mathbf{L_{TBS}}-\mathbf{H_2^*}||$.

\begin{lemma}[Distribution of $R_{\rm i2u}$ and $R_{\rm i2b}$]\label{lemma_RRprime}
Given the distance between the IoT cluster center and the serving UAV is $R_{\rm c2u}$, in the case of $R_{\rm c2u}>r_c$, where $r_c$ is the radius of IoT devices cluster, the PDF  of $R_{\rm i2u}$ is given by
\begin{align}
	f_{\rm R_{i2u}}(r) &= \frac{2r}{\pi r_c^2}\arccos\bigg(\frac{R_{\rm c2u}^2+r^2-r_c^2}{2 R_{\rm c2u} r}\bigg), \nonumber\\
	& {\rm if}\quad R_{\rm c2u}-r_c<r<R_{\rm c2u}+r_c,
\end{align}
otherwise, if $R_{\rm c2u}\leq r_c$, the PDF of $R_{\rm i2u}$ is
\begin{align}
	f_{\rm R_{i2u}}(r) &= \left\{\begin{aligned}
		&\frac{2r}{r_c^2},\quad 0<r<r_c-R_{\rm c2u} \\
		&\frac{2r}{\pi r_c^2}\arccos\bigg(\frac{R_{\rm c2u}^2+r^2-r_c^2}{2 R_{\rm c2u} r}\bigg),\\
		& \quad r_c-R_{\rm c2u}<r<R_{\rm c2u}+r_c,
	\end{aligned}\right.
\end{align}
The PDF of $R_{\rm i2b}$ follows a similar distribution by conditioning on $R_{\rm c2b}$, the distance between the IoT cluster center and the nearest TBS. Therefore, $f_{\rm R_{i2b}}(r)$ can be derived directly by replacing $R_{\rm c2u}$ by $R_{\rm c2b}$, and thus, omitted here.
\end{lemma}

As mentioned, we consider that the UAV delivers the data to the nearest TBS along the route. It means that assuming the reference UAV collecting the date from IoT devices and needs to travel to destination to deliver the package then back to source, the nearest TBS denotes the TBS which is the closest to UAV to $\mathbf{D}$ and $\mathbf{D}$ to $\mathbf{S}$ ($\mathbf{S}$-$\mathbf{D}$) segments, and $R_b$ refers to the distance between this TBS and the nearest point on the above two segments. 
\begin{theorem}[Distribution of $R_b$]\label{The_DisRb}
Here, we derive the CDF of $R_b$, which is a function of the length  and the angle of the segments,
\begin{align}
	F_{\rm R_b}(r,\theta,L_1,L_2) = 1-\exp\bigg(-\lambda_t {\rm Area}(r,\theta,L_1,L_2)\bigg),
\end{align}
where 
\begin{align}
	{\rm Area}(r,\theta,L_1,L_2) =& {\rm Area_1}(r,L_2)+{\rm Area_2}(r,\theta)+{\rm Area_3}(r,\theta,L_1)\nonumber\\
	&-{\rm Area_{hole}}(r),\nonumber\\
	{\rm Area_1}(r,L_2) =& \pi r^2+2L_2 r,\nonumber\\
	{\rm Area_2}(r,\theta) =& r^2\tan\bigg(\frac{\theta}{2}\bigg)-r^2,\nonumber
	\end{align}
\begin{align}
	&{\rm Area_3}(r,\theta,L_1) =\nonumber\\
	& \left\{\begin{aligned}
		&2r\bigg(L_1-\frac{r}{\sin(\theta)}\bigg)+\frac{r^2\pi}{2}, \quad 0\leq r\leq L_1\tan(\frac{\theta}{2}), \\
		&\frac{r^2(\theta_1-\sin\theta_1)}{2}+\bigg(L_1-r\tan\frac{\theta}{2}\bigg)\bigg(r\sin\frac{\theta_1}{2}\bigg)\sin\bigg(\frac{\theta_1}{2}\bigg),\\
		&\quad L_1\tan(\frac{\theta}{2})\leq r\leq L_1\tan(\theta),   \\
		&\frac{1}{2}\sin(\theta)xy+\theta_3 r^2-\frac{r^2\sin(2\theta_3)}{2}, \quad  L_1\tan(\theta)\leq r,
	\end{aligned}\right.\nonumber\\
	 &{\rm Area_{hole}}(r)= \nonumber\\
	& \left\{\begin{aligned}
		\frac{1}{2}\pi r^2+r r_t \cos(\theta_2)+\theta_2 r_t^2, \quad r < r_t\\
		\pi r_t^2, \quad r_t \leq r
	\end{aligned}\right.
\end{align}
in which,
\begin{align}
	\theta_1 &= \arcsin\bigg(\frac{l_1 \sin(\theta)-r}{r}\bigg)+\theta+\frac{\pi}{2},\nonumber\\
	\theta_2 &= \arcsin\bigg(\frac{r}{r_t}\bigg), \quad\theta_3 = \arcsin\bigg(\frac{z}{2r}\bigg),\nonumber\\
	x &= l_1-r\tan\bigg(\frac{\theta}{2}\bigg),\quad
	y = r\sin(\theta_4)+l_1\cos(\theta)+r\tan\bigg(\frac{\theta}{2}\bigg),\nonumber\\
	z &= \sqrt{(y-x\cos(\theta))^2+(x\sin(\theta))^2},\nonumber\\
	\theta_4 &= \frac{\pi}{2}-\arcsin\bigg(\frac{r-l_1\sin(\theta)}{r}\bigg),
\end{align}
all the related angles, lengths and geometry relations are shown in the proof.
\end{theorem}
\begin{IEEEproof}
See Appendix \ref{app_DisRb}.
\end{IEEEproof}

\section{Optimal Trajectory}
In this section, we mainly propose a multipurpose algorithm to solve the optimization problem $(\mathcal{P}_1)$ and $(\mathcal{P}_2)$. Recall that the objective of this work is data collection while delivering the package, we first maximize the size of collected data. If it is beyond the requirement of IoT devices, we minimize the round trip time while delivering all the data. To do so, we first need to characterize the time consumption when transferring the data. 

Observing that the time consumption defined in Definition \ref{Def_TimeCon} requires to take the expectation over SNR. Using the following lemma, we derive the PDF of SNR.
\begin{lemma}[PDF of SNR]
Coverage probability is the CCDF of SNR, given by
\begin{small}
\begin{align}\label{eq_Pcov}
	P_{\rm cov|R^{'}} &= \mathbb{P}\bigg(\frac{p(R)}{\sigma^2}>\gamma\bigg) =  \int_{r} \bigg(\sum_{k = 0}^{m_l}\frac{(m_l g_l(\sqrt{r^2+h^2})\gamma)^k}{k!}\nonumber\\
	&P_l(\sqrt{r^2+h^2})\exp(-m_l g_l(\sqrt{r^2+h^2}) \gamma)\nonumber\\
	&+\sum_{k = 0}^{m_n}\frac{(m_n g_n(\sqrt{r^2+h^2})\gamma)^k}{k!}\nonumber\\
	&P_n(\sqrt{r^2+h^2})\exp(-m_n g_n(\sqrt{r^2+h^2}) \gamma)\bigg)f_{\rm R}(r) {\rm d}r,\nonumber\\
P_{\rm cov|R_{u2b}} 	&= \mathbb{P}\bigg(\frac{p_u(R_{\rm u2b})}{\sigma^2}>\gamma\bigg) =  \bigg(\sum_{k = 0}^{m_l}\frac{(m_l g_l(\sqrt{R_{\rm u2b}^2+h^2})\gamma)^k}{k!}\nonumber\\
	&P_l(\sqrt{R_{\rm u2b}^2+h^2})\exp(-m_l g_l(\sqrt{R_{\rm u2b}^2+h^2}) \gamma)+\nonumber\\
	&\sum_{k = 0}^{m_n}\frac{(m_n g_n(\sqrt{R_{\rm u2b}^2+h^2})\gamma)^k}{k!}P_n(\sqrt{R_{\rm u2b}^2+h^2})\nonumber\\
	&\times\exp(-m_n g_n(\sqrt{R_{\rm u2b}^2+h^2}) \gamma)\bigg),
\end{align}
\end{small}
where $g_l(r) =  \gamma (\rho\eta_{\mathrm{l}})^{-1} r^{\alpha_{\mathrm{l}}}$ and $g_n(r) = \gamma (\rho\eta_{\mathrm{n}})^{-1} r^{\alpha_{\mathrm{n}}}$,  $\rho\in\{\rho_i,\rho_{u}\}$ depends on IoT or UAV communication and corresponding to $p\in\{p_{\rm i}(R_{\rm i2u}),p_{\rm b}(R_{\rm i2b})\}$, $R^{'}\in\{R_{\rm c2u},R_{\rm c2t}\}$ and $R\in\{R_{\rm i2u},R_{\rm i2b}\}$.
Hence, the PDF of SNR is derived by taking the first derivative of CCDF,
\begin{small}
\begin{align}\label{eq_fSNR}
	&f_{\rm SNR|R^{'}}(\gamma) 
	= \sum_{k = 1}^{m_l-1}\int_{r}\frac{(m_l g_l(\sqrt{r^2+h^2}))^k}{k!}P_l(\sqrt{r^2+h^2})\nonumber\\
	&\exp(-m_l g_l(\sqrt{r^2+h^2}) \gamma)f_{\rm R}(r)(m_l g_l(\sqrt{r^2+h^2})\gamma^{k}-k\gamma^{k-1}) {\rm d}r\nonumber\\
	&+\sum_{k = 1}^{m_n-1}\int_{r}\frac{(m_n g_n(\sqrt{r^2+h^2}))^k}{k!}P_n(\sqrt{r^2+h^2})\nonumber\\
	&\exp(-m_n g_n(\sqrt{r^2+h^2}) \gamma)f_{\rm R}(r)(m_n g_n(\sqrt{r^2+h^2})\gamma^{k}-k\gamma^{k-1}) {\rm d}r,\nonumber\\
	&f_{\rm SNR|R_{u2b}}(\gamma) 
	= \sum_{k = 1}^{m_l-1}\frac{(m_l g_l(\sqrt{R_{\rm u2b}^2+h^2}))^k}{k!}P_l(\sqrt{R_{\rm u2b}^2+h^2})\nonumber\\
	&\times\exp(-m_l g_l(\sqrt{r^2+h^2}) \gamma)(m_l g_l(\sqrt{R_{\rm u2b}^2+h^2})\gamma^{k}-k\gamma^{k-1}) \nonumber\\
	&+\sum_{k = 1}^{m_n-1}\frac{(m_n g_n(\sqrt{R_{\rm u2b}^2+h^2}))^k}{k!}P_n(\sqrt{R_{\rm u2b}^2+h^2})\nonumber\\
	&\exp(-m_n g_n(\sqrt{R_{\rm u2b}^2+h^2}) \gamma)(m_n g_n(\sqrt{R_{\rm u2b}^2+h^2})\gamma^{k}-k\gamma^{k-1}).
\end{align}
\end{small}
Note that the PDF of SNR of U2B channel is different from I2U and I2B channels owing to that $R_{\rm u2b}$ is not a random variable.
\end{lemma}
\begin{IEEEproof}
	The coverage probability equations is derived by the fact that (i)  $\Bar{F}_{\rm G}(g)=\frac{\Gamma_{u}(m,g)}{\Gamma(m)}$, where $\Gamma_{u}(m,g)=\int^{\infty}_{mg}t^{m-1}e^{-t}dt$ is the upper incomplete Gamma function, and (ii)  $\frac{\Gamma_{u}(m,g)}{\Gamma(m)}=\exp(-g)\sum^{m-1}_{k=0}\frac{g^{k}}{k!}$.
\end{IEEEproof}
Using the PDF of SNR derived above, we can now obtain the density of IoT clusters that need UAVs numerically,
\begin{align}
	\lambda_i^{'} = \lambda_i\exp(-\pi\lambda_t r_t^2),
\end{align}
where $r_t$ is the solution of $x$ to the following function,
\begin{align}
	\int_{0}^{\infty}\bar{C}_{\rm c2b}(x)f_{\rm SNR|R^{'}}(\gamma){\rm d}\gamma = c_t.
\end{align}

Recall that $M_{t|\mathbf{L_{IoT},L_{TBS}}}$ is the actual data UAV collected and delivered, which is smaller or equal to $M$. We now compute the time consumption of each link.

\begin{theorem}[Time Consumption]
	\label{theo_timecon}
	 Following Definition \ref{Def_TimeCon}, we take the expectation over SNR using the PDF derived above, the time consumption given the distance is 
	\begin{align}\label{eq_TR}
		T(R^{'}) &= \frac{M_{t|\mathbf{L_{IoT},L_{TBS}}}}{b_w}\int_{0}^{\infty}\frac{1}{\log_2(1+{\rm SNR|R^{'}})}f_{\rm SNR|R^{'}}(\gamma) {\rm d}\gamma,
	\end{align}
where $T(\cdot)\in\{T_{\rm c2u}(R_{\rm c2u}),T_{\rm u2b}(R_{\rm u2b}),T_{\rm c2b}(R_{\rm c2t})\}$.
\end{theorem}

\begin{figure}
	\centering
	\includegraphics[width= 1\columnwidth]{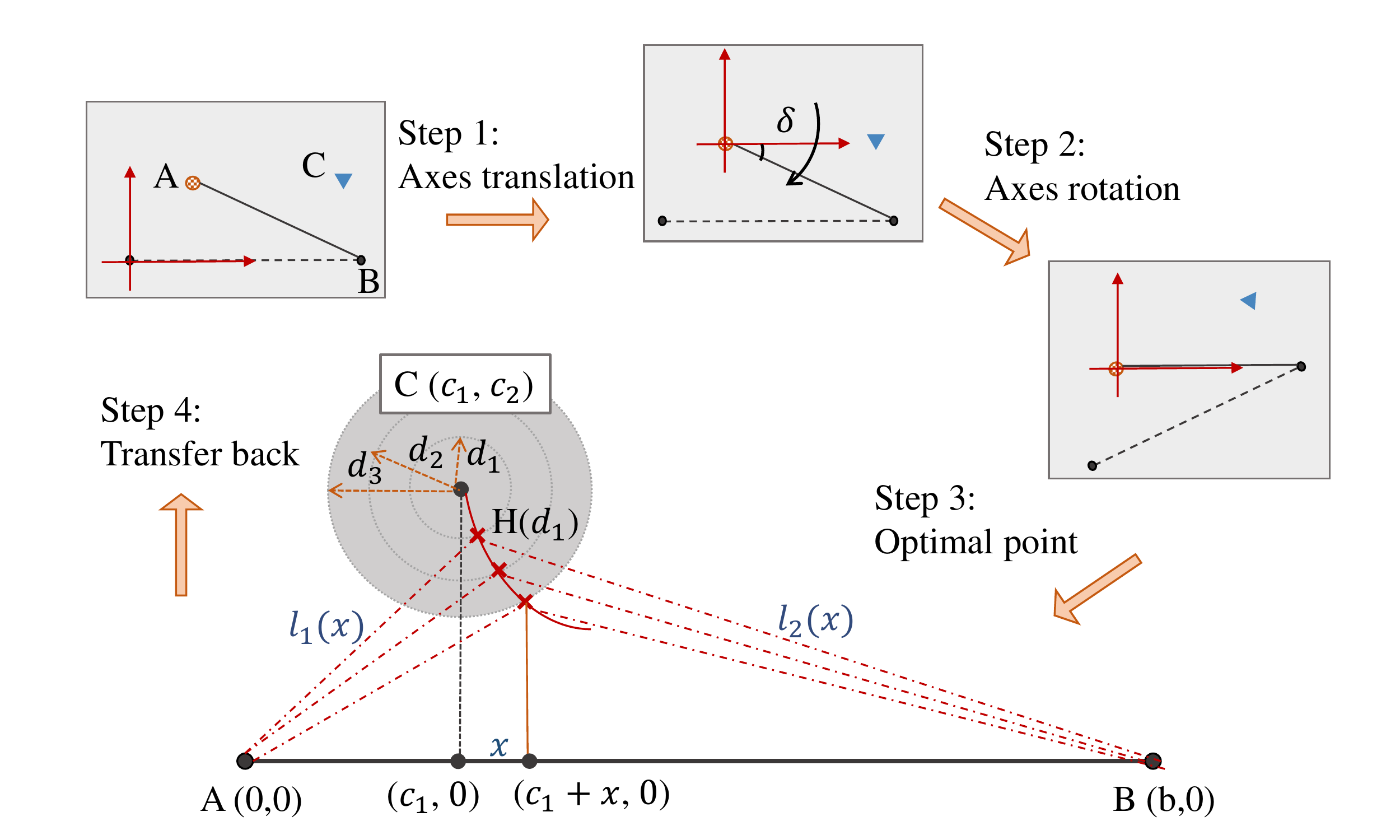}
	\caption{Illustration of finding the optimal hovering point.}
	\label{Fig_opthovering}
\end{figure}

We then minimize the UAV path which results in minimum traveling time and energy consumed during traveling. Suppose now the reference UAV travels from $\mathbf{A}$ to $\mathbf{B}$ while hovering at $\mathbf{H}(d)$ to communicate with $\mathbf{C}$ at $d$ away, and we aim to find the optimal $\mathbf{H}^*(d)$ that minimizes the length of the path:
\begin{align}
	\mathbf{H}^*(d) = \argmin_{|(\mathbf{H}(d)-\mathbf{C})|=d} |\mathbf{A}-\mathbf{H}(d)|+|\mathbf{H}(d)-\mathbf{B}|.
\end{align}
In what follows, we find the shortest path of UAV given $R_{\rm c2u}$ and $R_{\rm u2b}$.

\begin{lemma}[Shortest Path]\label{lemma_H1H2}
Let $\mathbf{Hov}_1(R_{\rm c2u})$ and $\mathbf{Hov}_2(R_{\rm u2b})$ be the hovering points which minimize the trajectory of UAV given the distances $R_{\rm c2u}$ and $R_{\rm u2b}$, respectively. The steps of obtaining $\mathbf{Hov}_1(R_{\rm c2u})$ and $\mathbf{Hov}_2(R_{\rm u2b})$ are shown in Fig.~\ref{Fig_opthovering}.
\end{lemma}
\begin{IEEEproof}
Our goal here is to find the optimal point $\mathbf{Hov_{\{1,2\}}}(d)$ which minimizes the length of the target trajectory given the final distance between $\mathbf{C}$ and UAV is $d$. We first establish a new coordinate system whose origin is $A$ and x-axis is $\vec{AB}$, as \textbf{\em Step 1} and \textbf{\em Step 2}, the relationship of these two coordinates are superposition of a rotation matrix and a translation vector:
\begin{align}
	\begin{bmatrix}
		x^{'}\\
		y^{'}
	\end{bmatrix}
 = 
 \begin{bmatrix}
\cos\delta & \sin\delta\\
-\sin\delta & \cos\delta
 \end{bmatrix}
	\begin{bmatrix}
	x\\
	y
\end{bmatrix}
-A ,
\end{align}
where $x^{'}$, $y^{'}$ are new coordinates and $x,y$ are the original coordinates, and $\delta$ is counter clockwise.

Based on the geometry relation shown in Fig. \ref{Fig_opthovering}, the shortest length can be derived numerically in \textbf{\em Step 3},
\begin{align}
l^{*} &= \min l(x) = l_1(x)+l_2(x)  \nonumber\\
&= \min \sqrt{(c_1+x)^2+(c_2-\sqrt{d-x^2})^2} \nonumber\\
&\quad+\sqrt{(b-c_1-x)^2+(c_2-\sqrt{d-x^2})^2},
\end{align}
above equation solved by setting $\frac{{\rm d}l(x)}{{\rm d}x}=0$ and $x^{*} = \argmin l(x)$.

Finally, transfer the coordination system back to the original one and we obtain the $\mathbf{Hov}_1(R_{\rm c2u})$ and $\mathbf{Hov}_2(R_{\rm u2b})$ given serving distance.
\end{IEEEproof}

 \begin{figure}
 	\centering
 	\includegraphics[width= 1\columnwidth]{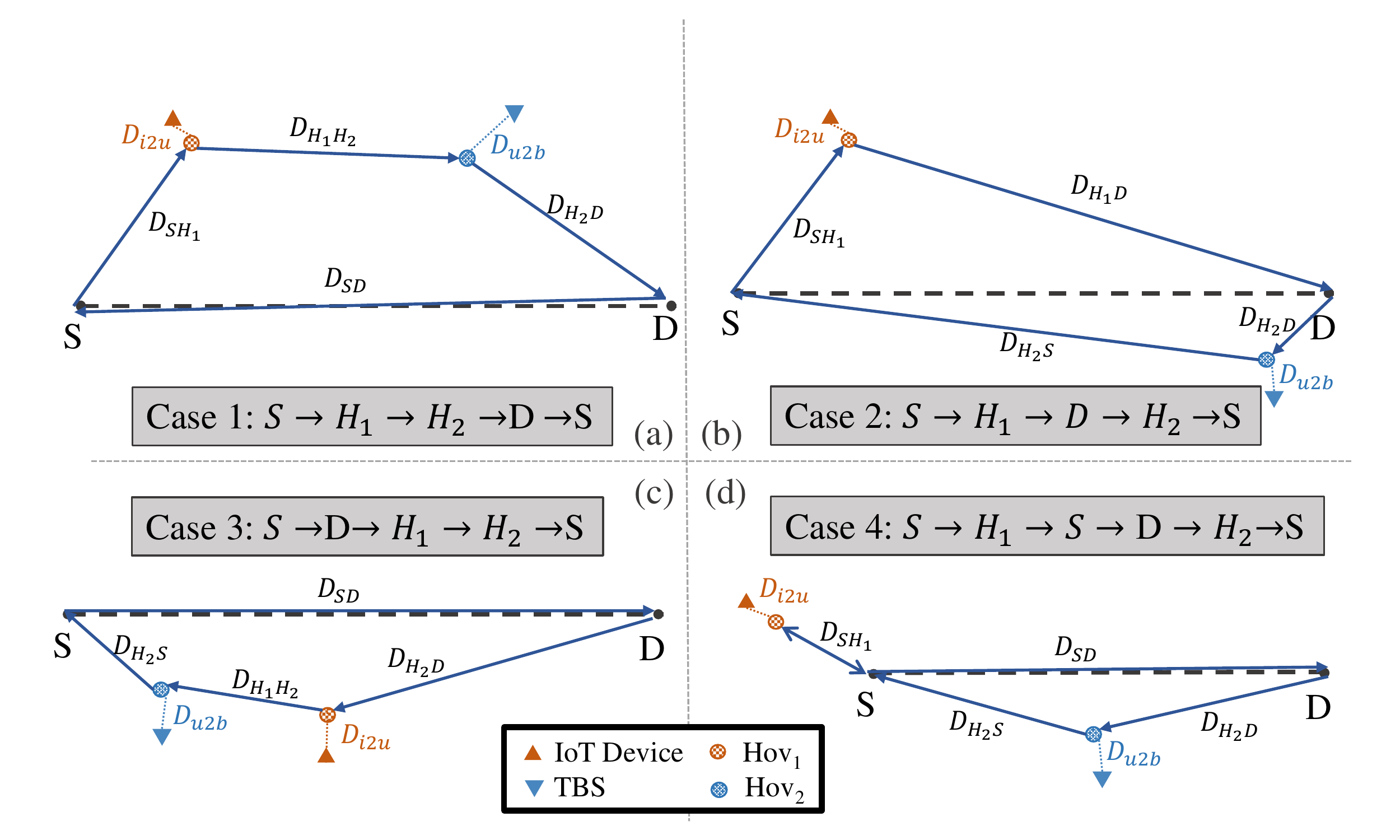}
 	\caption{Illustration of 4 possible trajectories of UAVs. Note that (i) $H_1$ and $H_2$ refer to hovering points $H_1(R_{\rm c2u})$ and $H_2(R_{\rm u2b})$, (ii) $\mathbf{Hov}_1(R_{\rm c2u})$ and $\mathbf{Hov}_2(R_{\rm u2b})$ are the optimal hovering points for each case, but not the final solutions to ($\mathcal{P}_1$) 	and ($\mathcal{P}_2$).}
 	\label{Fig_4cases}
 \end{figure}

 Now that we are ready for our algorithm and final solutions to the optimization problems. 
We divide the UAV trajectory into four possible routes, as shown in Fig. \ref{Fig_4cases}, and each of the total energy consumption and time needed are give below, where $D_{\{\cdot\}}$ are distances, and the subscript $H_1$ and $H_2$ refer to hovering points $\mathbf{Hov}_1(R_{\rm c2u})$ and $\mathbf{Hov}_2(R_{\rm u2b})$.
 
 \textbf{Route 1: $S\to IoT\to TBS\to D\to S$.} (The reference UAV starts at $\textbf{S}$, collects and delivers the data and then delivers the package and finally back to $\textbf{S}$.)
 
 \begin{align}
 	T_1 = & \frac{D_{\rm SH_1}+D_{\rm H_{1}H_{2}}+D_{  \rm H_{2}D}}{v_p} \nonumber\\
 	&+T_{\rm c2u}(R_{\rm c2u})+T_{\rm u2b}(R_{\rm u2b})+\frac{D_{\rm SD}}{v},\nonumber\\
 	E_{t1} =& \frac{D_{\rm SH_1}+D_{\rm H_{1}H_{2}}+D_{\rm H_{2}D}}{v_p}p_{mp} \nonumber\\
 	&+(T_{\rm c2u}(R_{\rm c2u})+T_{\rm u2b}(R_{\rm u2b}))p_{sp}+\frac{D_{\rm SD}}{v}p_{m}.
 \end{align}

 \textbf{Route 2: $S\to IoT\to D\to TBS\to S$.} (The reference UAV starts at $\textbf{S}$, collects the data and delivers the package, then delivers the data and finally back to $\textbf{S}$.)
 
 \begin{align}
 	T_2 =&  \frac{D_{\rm SH_1}+D_{\rm H_{1}D}}{v_p}+T_{\rm c2u}(R_{\rm c2u})+T_{\rm u2b}(R_{\rm u2b}) \nonumber\\
 	&+\frac{D_{\rm H_{2}D}+D_{\rm H_{2}S}}{v},\nonumber\\
 	E_{t2} =& \frac{D_{\rm SH_1}+D_{\rm H_{1}D}}{v_p}p_{mp}+T_{\rm c2u}(R_{\rm c2u})p_{sp}+T_{\rm u2b}(R_{\rm u2b})p_{s} \nonumber\\
 	&+\frac{D_{\rm H_{2}D}+D_{\rm H_{2}S}}{v}p_m.
 \end{align}
 
 \textbf{Route 3: $S\to D\to IoT\to TBS\to S$} (The reference UAV starts at $\textbf{S}$, delivers the package first, then collects and delivers the data and finally back to $\textbf{S}$.)
 
 \begin{align}
 	T_3 &=  \frac{D_{\rm SH_2}+D_{\rm H_{1}H_{2}}+D_{\rm DH_1}}{v} \nonumber\\
 	&+T_{\rm c2u}(R_{\rm c2u})+T_{\rm u2b}(R_{\rm u2b})+\frac{D_{\rm SD}}{v_p},\nonumber\\
 	E_{t3} &= \frac{D_{\rm SH_2}+D_{\rm H_{1}H_{2}}+D_{\rm DH_1}}{v}p_m \nonumber\\
 	&+(T_{\rm c2u}(R_{\rm c2u})+T_{\rm u2b}(R_{\rm u2b}))p_s+\frac{D_{\rm SD}}{v_p}p_{mp}.
 \end{align}
 
 \textbf{Route 4: $S\to IoT\to S\to D\to TBS\to S$} (The reference UAV starts at $\textbf{S}$, collects the data first, then back to $\textbf{S}$ to take the package and delivers the data and package.)
 
 \begin{align}
 	T_4 &=  \frac{2D_{\rm SH_1}+D_{\rm SH_{2}}+D_{\rm H_{2}D}}{v}+T_{\rm c2u}(R_{\rm c2u}) \nonumber\\
 	&+T_{\rm u2b}(R_{\rm u2b})+\frac{D_{\rm SD}}{v_p},\nonumber\\
 	E_{t4} &= \frac{2D_{\rm SH_1}+D_{\rm SH_{2}}+D_{\rm H_{2}D}}{v}p_m \nonumber\\
 	&+(T_{\rm c2u}(R_{\rm c2u})+T_{\rm u2b}(R_{\rm u2b}))p_s+\frac{D_{\rm SD}}{v_p}p_{mp}.
 \end{align}

In the following text, we propose an algorithm that solves the optimization problems given $\mathbf{L_{IoT}}$ and $\mathbf{L_{TBS}}$, as shown in Algorithm \ref{Alg_OptRoute}. The detailed steps are provided in the following text.

We first find the distance between $\mathbf{L_{IoT}}$ to nearest point on $\mathbf{S}$-$\mathbf{D}$ segment is $R_{\rm max}$. Our algorithm start from conditioning on each $R_{\rm c2u} \in (0,R_{\rm max})$. For each $R_{\rm c2u}$, we obtain $\mathbf{Hov_1}$ using Lemma \ref{lemma_H1H2}.

\textbf{\em Variables and notations.} $E_{\rm con}$, $T_{\rm Mmax}$, $T_{\rm total}$ and $M_{\rm max}$ are the temporary variables used in iterations. Let $\Phi_{\rm Hov_1}$ be the point set of $\mathbf{Hov_1}$, and let $\Phi_{\rm Hov_2|Hov_1}$ be the  point set of $\mathbf{Hov_2}$ given $\mathbf{Hov_1}$. In addition, this algorithm solves the conditional optimization problem: all the outputs are actually conditioned the locations of IoT cluster and TBS. However, to simplify the notation, here we use $M_t$ instead of $M_{\rm t|\mathbf{L_{IoT},L_{TBS}}}$ and the same applies to $E_{\rm total}$ and $T$. As mentioned, we consider the transmitted data over bandwidth as one parameter, therefore, $M_{\{\cdot\}}$  refers to $\frac{M_{\{\cdot\}}}{b_w}$.

\textbf{\em Part1: maximum transferred data.} We first check the maximum date $M_{\rm max}$ that UAV can transfer for given $R_{\rm c2u}$. 
Notice that the time consumption of collecting data $T_{\rm c2u}$ and $T_{\rm u2b}$ are only functions of the distances $R_{\rm c2u}$ and $R_{\rm u2b}$ (assume that $\frac{M}{b_w}$ and $\rho_i$ are predetermined) and $\mathbf{Hov_1}$ can be obtained by using Lemma \ref{lemma_H1H2}. Therefore, the optimal $R_{\rm u2b}$ and corresponding $\mathbf{Hov_2}$ that maximize $M_{\rm max}$ given $R_{\rm u2b}$ can be easily obtained. $R_{\rm u2b}$ is derived by solving 
\begin{align}
	\frac{M_{\rm max}}{b_w} = \frac{(B_{\rm max}-\frac{l_p}{v_p}p_{mp}-\frac{l}{v}p_{m})}{T_{\rm c2u}(R_{\rm c2u})p_{\{s,sp\}}+T_{\rm u2b}(R_{\rm u2b})p_{\{s,sp\}}}, 
\end{align}
where $l$ and $l_p$ are the traveling distances with/without package, and the denominator $p_{sp}$ or $p_{s}$ is owing to the states of UAVs when communicating: with or without package, and  $\mathbf{Hov_2}$ is obtained by Lemma \ref{lemma_H1H2}.

\textbf{\em Part2: outputs update.} We then compare $M_{\rm max}$ with $M$. If $M_{\rm max} < M$, which means that the UAV is unable to collect and transmit all the data. In this case, we update $M_t$ as well as all the outputs, only if $M_{\rm max}$ is greater than $M_{\rm t}$. ($M_t$ saves the largest value of $M_{\rm max}$ in previous loops if none of them are greater than $M$, otherwise, it equals to $M$.)
If $M_{\rm max} > M$, which means that the UAV is able to transfer all the data from IoT devices to the TBS within the available time and energy. In this case, we first obtain the subset $\Phi_{\rm Hov_2|Hov_1}^{'}$ of $\Phi_{\rm Hov_2|Hov_1}$ which is the set of $\mathbf{Hov_2}$ that can transfer all $M$. Within $\Phi_{\rm Hov_2|Hov_1}^{'}$, we then find the optimal $\mathbf{Hov_2}^{(2)}$ which minimizes the round trip time $T_{\rm total}$ and  outputs update if $T_{\rm total}$ is less than previous value.

\begin{algorithm}
	\caption{Algorithm for UAV Optimal Route Planning}
	\SetAlgoLined
	\DontPrintSemicolon
	\KwIn{${M}$: Required transmitted data\newline
		$({\rm IoT_x,IoT_y})$: Location of IoT cluster center\newline
		$({\rm TBS_x,TBS_y})$: Location of TBS}  
	\KwOut{${\rm T}$: Minimum time to finish the assisgnment \newline
		${\rm H_{1}^{*} ,H_{2}^{*} }$: Locations of UAV hovering to provide service\newline
		${\rm M_{t}}$: Maximum collected/delivered data \newline
		${\rm E_{\rm total}}$: Energy consumption of UAV\newline
		${\rm Route}$: Route of UAVs}
	\textbf{Initialization:} ${\rm T} = \infty$, $E_{\rm total} = \infty$, ${\rm M_t} = 0$, ${\rm H_{1}^{*}} = \emptyset$, ${\rm H_{2}^{*}} = \emptyset$, ${\rm Route} = 0$\newline
	\SetKwFunction{FMain}{OptimalRoute}
	\SetKwProg{Fn}{Function}{:}{}
	\Fn{\FMain{$M,({\rm IoT_x,IoT_y}),({\rm TBS_x,TBS_y})$}}{
		{\rm Step 1}: Find the distance $R_{\rm max}$ from $({\rm IoT_x,IoT_y})$ to the closest point on S-D segment.\newline
		$R_{\rm u2b}$ given $k$-th iteration is $k\times step$, where $step$ is the iteration step and $k_{\rm max}$ = $\frac{R_{\rm max}}{step}$. \newline
		\ForEach{$ k \leq k_{\rm max} $}{
			Within the limited energy, find $\mathbf{Hov_2}^{(1)}$ among 4 possible UAV routes which maximizes the transmitted data ${\rm M_{max}}$.\newline
			Go to the next loop if ${\rm M_{max}} < 0$.\newline
			{\eIf{$ M_{\rm max} \leq M $}
				{Update outputs if ${\rm M_{max}}$ is larger than previous iterations. }
				{Find $\mathbf{Hov_2}^{(2)}$ which minimizes the total ${\rm T_{total}}$ while transmitting all the data within the limited energy.\newline
					If $M_t < M$, outputs update and $M_t = M$. \newline
					If $M_t = M$, outputs update if $T_{\rm total}$ is shorter than previous iterations.
				}
			}
		}
		\textbf{return} ${\rm T, H_{1}^{*} ,H_{2}^{*} , M_t, E_{\rm total}, Route}$ 
	}
	\textbf{End Function}
	\label{Alg_OptRoute}
\end{algorithm}
Before we process the next step to obtain the average round trip time, energy consumption, and maximum transmitted data. We would like to clarify the goal of the above algorithm is to obtain the optimal trajectory: for a given location of the IoT cluster and TBS, if the UAV can deliver all the data, we minimize the required time; if the UAV cannot, we maximize the collected/delivered data. This optimal trajectory can be either Route 1, 2, 3, or 4, which depends on the locations.

\begin{figure}
	\centering
	\includegraphics[width= 1\columnwidth]{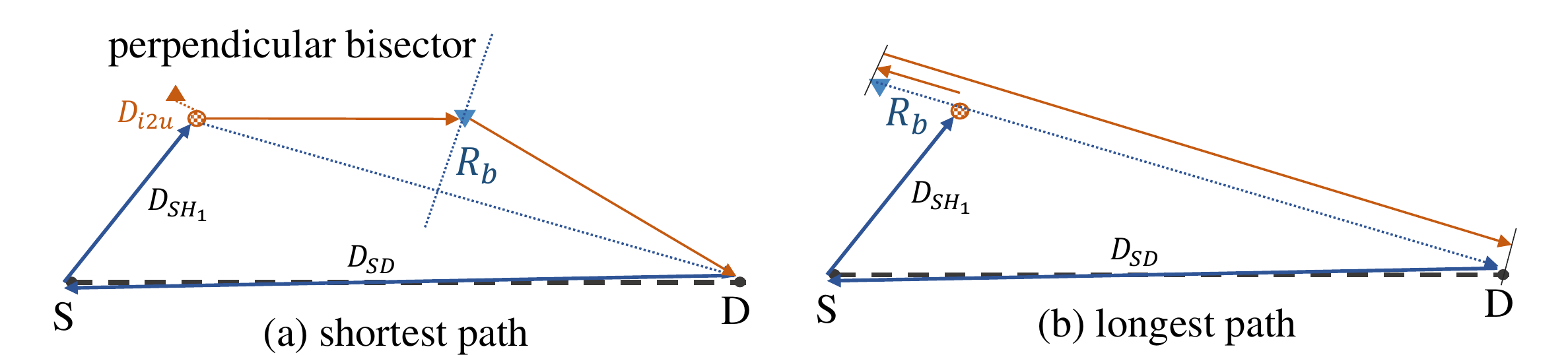}
	\caption{Proof for the upper bound. \textbf{(a)} the shortest path: when TBS is located at the perpendicular bisector, and \textbf{(b)} the longest path: when TBS is located at the opposite point on the extension of $\mathbf{L_{IoT}}$-$\mathbf{D}$ segment.}
	\label{proof_shortlong}
\end{figure}

The above algorithm conditions on $\mathbf{L_{IoT}}$ and $\mathbf{L_{TBS}}$ which are random variables in our system, and we are interested in the power consumption, minimum round trip time, and collected/delivered data in a general case and the average system performance. However, it is difficult to obtain the joint PDF of the locations of the IoT cluster and the nearest TBS, especially when they are correlated; hence, the method in Definition \ref{Def_generalcase} is hard to compute. Alternatively, we use the method provided below, which is an upper bound of the system's performance.
 \begin{theorem}[General Case]\label{Def_montecarlo}
 We first take the integral over $\mathbf{L_{TBS}}$.  Since the UAV goes to the nearest TBS along the route, hence, we should take the integral over $R_b$, which is
 \begin{align}
 	&\{T|\mathbf{L_{IoT}},E_{\rm total|\mathbf{L_{IoT}}},\frac{M_{\rm t|\mathbf{L_{IoT}}}}{b_w}\} =  \int_{0}^{\infty}\biggl\{T_{\rm min|\mathbf{IoT,TBS}},\nonumber\\
 	&E_{\rm total|\mathbf{L_{IoT},L_{TBS}}} ,\frac{M_{\rm t|\mathbf{L_{IoT},L_{TBS}}}}{b_w}\biggr\} F_{\rm Rb}(r,\theta,l_1,L_2){\rm d}r, \label{eq_intTBS}
 \end{align}
 where $l_1 = ||\mathbf{L_{IoT}}-\mathbf{D}||$ and $\theta = \arccos\bigg(\frac{L_2-\mathbf{L_{IoT}}_x}{||\mathbf{L_{IoT}}-\mathbf{D}||}\bigg)$. 
 
  We then use the same method and take the integral over the location of $\mathbf{L_{IoT}}$. While we admit that UAVs can provide the service to any IoT clusters which are far from TBSs, here we only consider the nearest one, the nearest to $\mathbf{S}$-$\mathbf{D}$ pair,
 \begin{align}
 	\{T,&E_{\rm total},\frac{M_{\rm t}}{b_w}\} =\nonumber\\
 	& \int_{0}^{\infty}\{T|\mathbf{L_{IoT}},E_{\rm total|\mathbf{L_{IoT}}} ,\frac{M_{\rm t|\mathbf{L_{IoT}}}}{b_w}\}F_{\rm IoT}(r){\rm d}r, \label{eq_intIoT}
 \end{align}
 where,
 \begin{align}
 	F_{\rm IoT}(r) = F_{\rm Rb}(r,0,0,L_2),
 \end{align}
and replace $\lambda_t$ by $\lambda_i^{'}$.
\end{theorem}
\begin{algorithm}
	\caption{Algorithm for Monte-Carlo Integration}
	\SetAlgoLined
	\DontPrintSemicolon
	\KwIn{$r$: Integral variable}  
	\KwOut{${\rm T}(r)$: Round trip time\newline
		${\rm M}(r)$: Collected/delivered data\newline
		${\rm E}(r)$: Energy consumption}
	\SetKwFunction{FMain}{MonteCarloIntegration}
	\SetKwProg{Fn}{Function}{:}{}
	\Fn{\FMain{$r$}}{
		Generate a vector for the values of $R_b$: $X_{\rm R_{b,1}},\cdots,X_{\rm R_{b,n}}$ \newline
		Corresponding CDF ($\mathbb{P}(R_b < X_{\rm R_{b,i}})$): $F_{\rm X_{R_{b,1}}},\cdots,F_{\rm X_{R_{b,n}}}$ \newline
		\ForEach{$ k \leq {\rm iteration}$}{
			Generate $U \sim {\rm Uniform}(0,1)$\newline
			Find the nearest point in $F_{\rm X_{R_{b,i}}}$\newline
			Find corresponding $X_{\rm R_{b,i}}$\newline
			Using Algorithm 1, find the optimal ${\rm Tmin}$, $E_{\rm total}$ and $M_t$
		}
		Find the mean: $T(r) = {\rm mean}(Tmin)$, $E(r) = {\rm mean}(E_{\rm total})$,  $M(r) = {\rm mean}(M_t)$
		\textbf{return} $T(r),M(r),E(r)$ 
	}
	\textbf{End Function}
	\label{Alg_MCIntegral}
\end{algorithm}
\begin{remark}\label{Rem_upperbound}
	Observing that the conditional system parameters in (\ref{eq_intTBS}) and (\ref{eq_intIoT}) are the outputs of $(\mathcal{P}_{1})$ and $(\mathcal{P}_{2})$ and functions of the integral variable $r$  which cannot be obtained in closed form. Therefore,
to solve the integration in Theorem \ref{Def_montecarlo}, Monte-Carlo integration method is used and we provide an algorithm to solve the above integration, as shown in Algorithm \ref{Alg_MCIntegral}. In addition, we only know the probability of the distance to the nearest TBS and IoT cluster center instead of the exact location (e.g., the probability of $R_b$ instead of $\mathbf{L_{TBS}}$). Hence, in our codes, we use the upper bound: the longest route (e.g., taking route 1, for example, UAV travels from IoT cluster to TBS then to destination, and the distance between  $\mathbf{L_{IoT}}$-$\mathbf{D}$ segment and TBS is $R_b$. The shortest path for UAV is when TBS is located at the perpendicular bisector, and the longest path is when TBS is located at the opposite point on the extension of $\mathbf{L_{IoT}}$-$\mathbf{D}$ segment), as shown in Fig. \ref{proof_shortlong}.
\end{remark}

 \vspace{-4mm}

\section{Numerical Results}
	\begin{table}\caption{Table of Parameters}\label{par_val}
	\centering
	\begin{center}
		\resizebox{1\columnwidth}{!}{
			\renewcommand{\arraystretch}{1}
			\begin{tabular}{ {c} | {c} | {c}  }
				\hline
				\hline
				\textbf{Parameter} & \textbf{Symbol} & \textbf{Simulation Value}  \\ \hline
				Density of TBS and IoT cluster & $\lambda_{\rm t}$,$\lambda_{\rm i}$ & 1 km$^{-2}$ \\ \hline
				IoT cluster radius & $r_c$ & 50  m\\ \hline
				Achievable rate threshold & $c_t $ & $1$ bps/Hz\\\hline
				Average package weitht & $\bar{w}$ & $1$ kg \\\hline
				Optimal with/without package velocity & $v_p,v$ & 12.4, 10 m/s\\ \hline
				Serving-related power (with/without package) & $p_s,p_{sp}$ & 252 J, 178 J\\ \hline
				Traveling-related power (with/without package) & $p_m,p_{mp}$ & 193 J, 159 J \\ \hline
				UAV altitude & $h$ & 100 m\\\hline
				Battery capacity & $B_{\rm max}$ & 177.6 W$\cdot$H \\\hline
				N/LoS environment variable & $a, b$ & 4.9 0.43 \\\hline
				Transmission power & $\rho_{\rm i},\rho_{\rm u}$ & 0.1 mW, 0.1 W\\\hline
				Noise power & $\sigma^2 $ & $10^{-9}$ W\\\hline
				N/LoS and TBS path-loss exponent & $\alpha_{\rm n},\alpha_{\rm l},\alpha_{\rm t}$ & $4,2.1,4$ \\\hline
				N/LoS fading gain & $m_{\rm n},m_{\rm l}$ & $1,3$ \\\hline
				N/LoS additional loss& $\eta_{\rm n},\eta_{\rm l}$ & $-20,0$ dB 
				\\\hline\hline
		\end{tabular}}
	\end{center}
\end{table}

In this section, we validate our analytical results with simulations and evaluate the impact of various system parameters on the network performance. Unless stated otherwise, we use the simulation parameters as listed herein Table \ref{par_val}.

\begin{figure}
	\centering
	\includegraphics[width= 0.99\columnwidth]{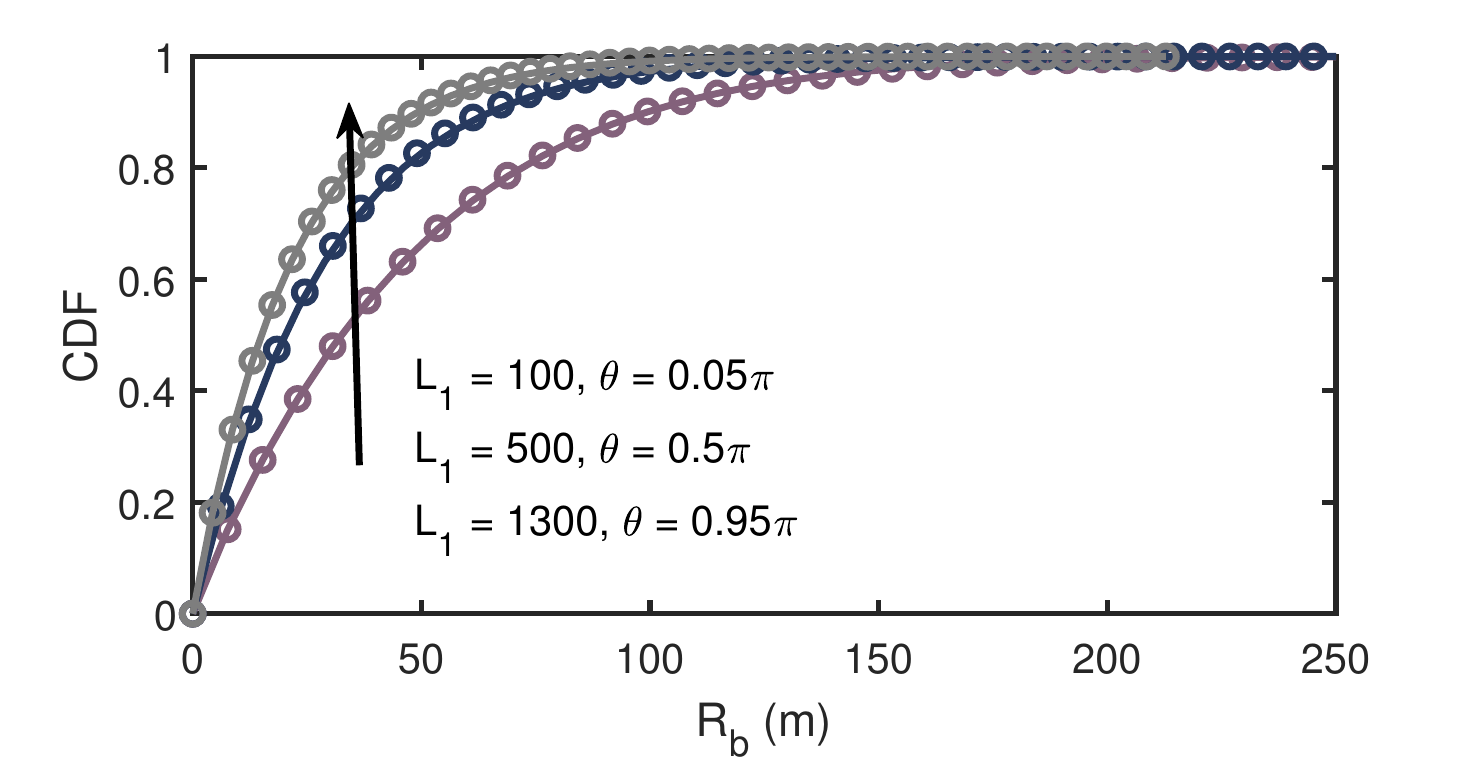}
	\caption{Simulation results vs analysis results of the distribution of $R_b$, conditioned on $L_2 = 1000$ m.}
	\label{Res_SimRb}
\end{figure}

\begin{figure*}
	\centering
	\includegraphics[width= 1.8\columnwidth]{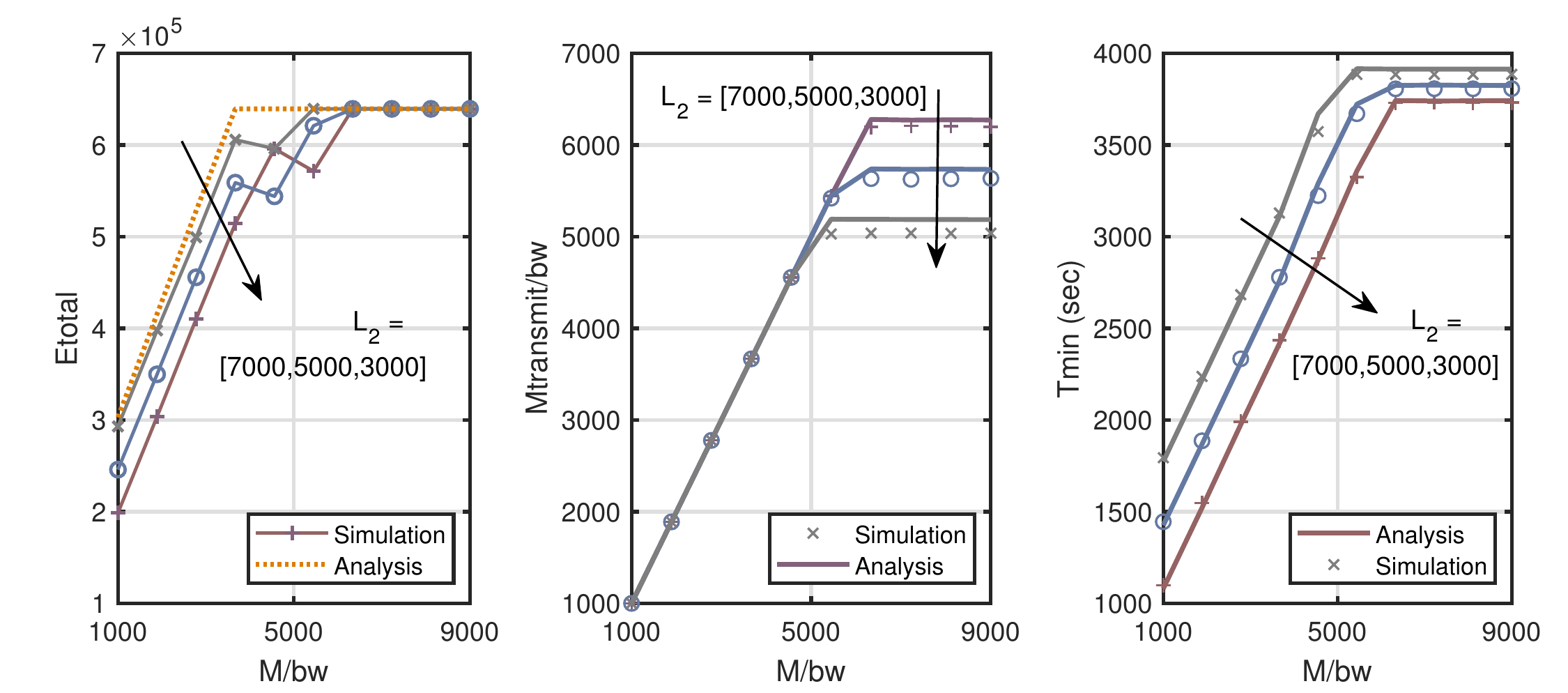}
	\caption{Simulation and analysis results of total energy consumption, maximum collected/delivered data  and minimum round trip time vs required data from IoT devices side under different $\mathbf{S}$-$\mathbf{D}$ distances. Note that we only plot one analysis result ($L_2 = 7000$ km) in energy consumption figure.}
	\label{Res_difMbw}
\end{figure*}

In our simulation, we first generate the locations of IoT cluster centers, TBSs and the locations of IoT devices. For each realization, we use the Algorithm \ref{Alg_OptRoute} to obtain the optimal route, as mentioned it can be either Route 1, 2, 3, or 4, which depends on the locations. We also obtain the maximum transferred data, round trip time, package delivery time, and the energy consumption of the optimal trajectory. Finally, we run a large number of iterations to ensure accuracy and obtain the average performance.

\begin{figure*}
	\centering
	\includegraphics[width= 1.8\columnwidth]{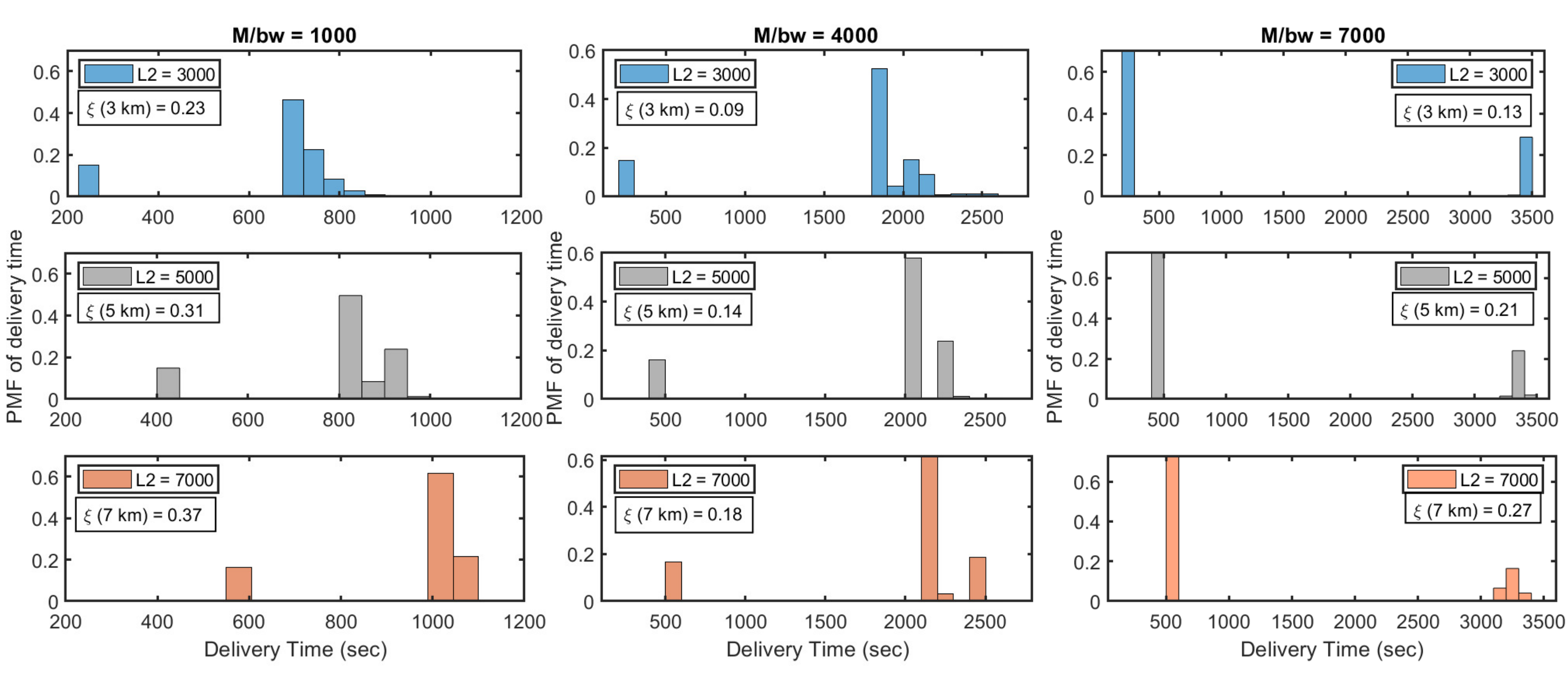}
	\caption{Probability of delivery time and delivery efficiency under different $L_2$ and required amount of data.}
	\label{Res_hisgram}
\end{figure*}
For the simulation of the considered system setup, we first compute the accuracy of the distribution of $R_b$, as derived in Theorem \ref{The_DisRb}. For given $L_2 = 1$ km, we plot the CDF of $R_b$ under 3 different $L_1$ and $\theta$, as shown in Fig.~\ref{Res_SimRb}.
We then use this distribution in final Monte-Carlo integration of $R_b$ and $\mathbf{L_{IoT}}$.

In Fig. \ref{Res_difMbw}, we plot the average total energy consumption, average maximum collected/delivered data, and the average  minimum round trip time. We first clarify that the gaps between simulation and analysis results are due to the fact that we use the upper bound in Monte-Carlo integration, as mentioned in Remark \ref{Rem_upperbound}.   Besides, we only plot one analysis curve ($L_2 = 7$ km, dash line). This is because the analysis results are the upper bound of the simulation results and the same trend applies to all distances. All the analysis results are slightly higher than the simulation results.

The maximum collected/delivered data and minimum round trip time, as expected, increase with the increasing of required data until maximum achievable value, which is limited by the onboard battery of UAVs. Total energy consumption, however, shows a different trend. This is because our goal in this work is to maximize the delivered data and minimize the round trip time within the limited energy of UAVs instead of minimizing the energy consumption of UAVs. Moreover, the minimum time path is different from the minimum energy consumption path and the maximum collected/delivered data path. This is because of the different optimal velocities and power consumption of UAVs with/without the package. Observing that UAVs with the package have higher velocity but higher energy consumption. In this case, the required transmitted data size is low and the UAV prefers to choose the trajectory that has minimal distance even if it results in higher energy consumption. However, if the transmitted data size is large, the UAV has to deliver the package first to save energy to deliver data.

To further illustrate the trend of total energy consumption, we take route 1 and route 4, for example. From the perspective of time, route 1 seems better since UAVs collect and deliver the data on the way of delivering package. However, on the side of energy consumption, route 4 may result in less traveling, and hovering energy since the energy consumption of UAVs is very sensitive to the total weight. Therefore, traveling without a package can save a large amount of energy. In the case when IoT devices require to transfer a large amount of data, UAVs prefer route 4 owing to having more energy left to collect and deliver the data.

Fig. \ref{Res_hisgram} shows the distribution of package delivery time, and delivery efficiency under different $\mathbf{S}$-$\mathbf{D}$ distances and required transmitted data sizes. Interestingly, $L_2 = 3$ km achieves the lowest delivery efficiency with the shortest delivery time. The results can be explained as (i) delivery time $T_{\rm nodata}$ in the case of $L_2$ is the shortest, hence, the denominator of $\xi$ (3 km) is the lowest, (ii) when $L_2$ is short, the nearest IoT cluster is likely to be located further away, the same as the nearest TBS, which means that UAVs need to fly for a larger circle to transfer data.

Observing that the delivery efficiency first decreases with the increased required collected/delivered data and then decreases. The same reason as the trend of energy consumption: when the requirement collected/delivered data size is large, UAVs prefer to deliver the package first in this case they can have more energy left to collect and transfer the data. 	That is, with the increase of the required transferred data, UAVs spend a longer time on data transmission; with further increase of the required transferred data, UAVs deliver the package first to save energy to transfer more data.

Another attractive phenomenon in Fig. \ref{Res_hisgram} is the delivery time gap. This is because the transmission time is relatively much longer than the delivery time. The low value of delivery time (left-hand side bars) is caused by the UAV delivering the package first, hence, the delivery time is only composed of the traveling time; and the high value of delivery time (right-hand side bars) is caused by the UAV transferring data (collecting data from IoT devices), hence, the delivery time is composed of communication time and traveling time.

Before concluding this section, we also discuss why not just deliver the package first and then deliver the data. To answer this question, we provide the following comparison between the energy consumption, round trip time, and maximum transferred data, from the perspective of operators, in Table \ref{par_com}. These four points are obtained given the required transmission data are M/bw $= [1,4,6,10]\times 10^3$ bit/Hz and $L_2 = 5$ km. When the required collected/delivered data size is small, delivering the package first can save UAVs' energy while the round trip time is slightly longer. However, when the required collected/delivered data is large, the maximum possible delivered data of delivering the package first is lower and the total round trip time is longer, compared to the optimal trajectory. The results in Table \ref{par_com} reveal a trade-off between higher quality of delivery package and higher time and energy efficiency. If the operators hope to deliver more packages, then the optimal trajectory is better, however, the relative package delivery time is longer.
	\begin{table}\caption{Comparison ($L_2$ = 5 km)}\label{par_com}
		\vspace{-8mm}
	\centering
	\begin{center}
		\resizebox{1\columnwidth}{!}{
			\renewcommand{\arraystretch}{1}
			\begin{tabular}{ {c} | {c}  }
				\hline
				\hline
				\textbf{Method} & \textbf{Energy Consumption} $(10^5)$ J \\ \hline
				Deliver package first & (2.39, 4.76, 6.4, 6.4)\\ \hline
				Optimal trajectory & (2.45, 5.59, 6.4, 6.4)
				\\\hline
				\textbf{Method} &  \textbf{Max Data Transferred} $(10^3)$ bits/Hz\\ \hline
				Deliver package first & (1, 3.7, 5.4, 5.4) \\ \hline
				Optimal trajectory & (1, 3.7, 5.6, 5.6) 
				\\\hline
               \textbf{Method} & \textbf{Total Time} $(10^3)$ sec \\ \hline
				Deliver package first & (1.6, 2.9, 3.9, 3.9) \\ \hline
				Optimal trajectory & (1.4, 2.7, 3.8, 3.8)
				\\\hline
				\hline
		\end{tabular}}
	\end{center}
\end{table}

\section{Conclusion}
This paper presented a novel system model in which UAVs simultaneously perform multiple tasks: data collection/delivery and package delivery. We investigated the possibility for UAVs to deliver the data and packages at the same time, and provided an algorithm to maximize the possible collected/delivered data while minimizing the round trip time. Moreover, our system model and results are applied to different scenarios since all the locations are random.

This work tapped a new aspect of the applications of UAVs. Instead of dedicated UAVs, multipurpose drones seem more efficient and realistic in real life. While UAVs are widely used in last-mile deliveries, they can also be used in wireless communication networks to fully display their benefits: flexibility, capability to optimize their locations in real-time, and providing an additional capacity of the cellular networks.
	\appendix
	\subsection{Proof of Theorem \ref{The_DisRb}}\label{app_DisRb}
\begin{figure}[ht]
	\label{proof_Rb}
	\centering
	\includegraphics[width= 0.7\columnwidth]{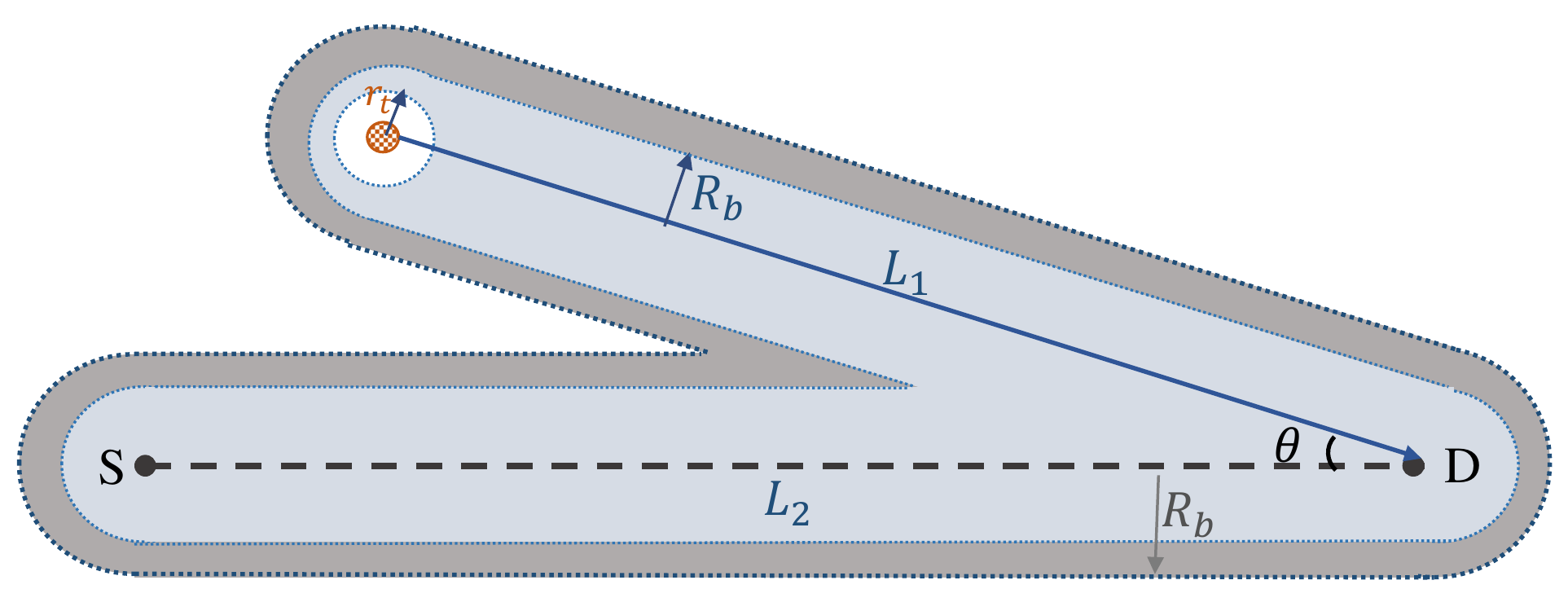}
	\caption{Illustrate of distance distribution of $R_b$.}
\end{figure}

In this section, we provide the proof of the distribution of $R_b$. The same method as computing the first nearest neighbor in PPP, the probability of $R_b > r$ equals to the probability of none point falling in the certain area, say $A$,
\begin{align}
\mathbb{P}(R_b > r) = \mathbb{P}(\mathcal{N}(A) = 0),	
\end{align}
where $A$ is the shadowing area in Fig.~\ref{proof_Rb}. In this case, to compute the CDF of $R_b$, we need to compute the area of A.

\begin{figure}[ht]
		\centering
		\includegraphics[width= 1\columnwidth]{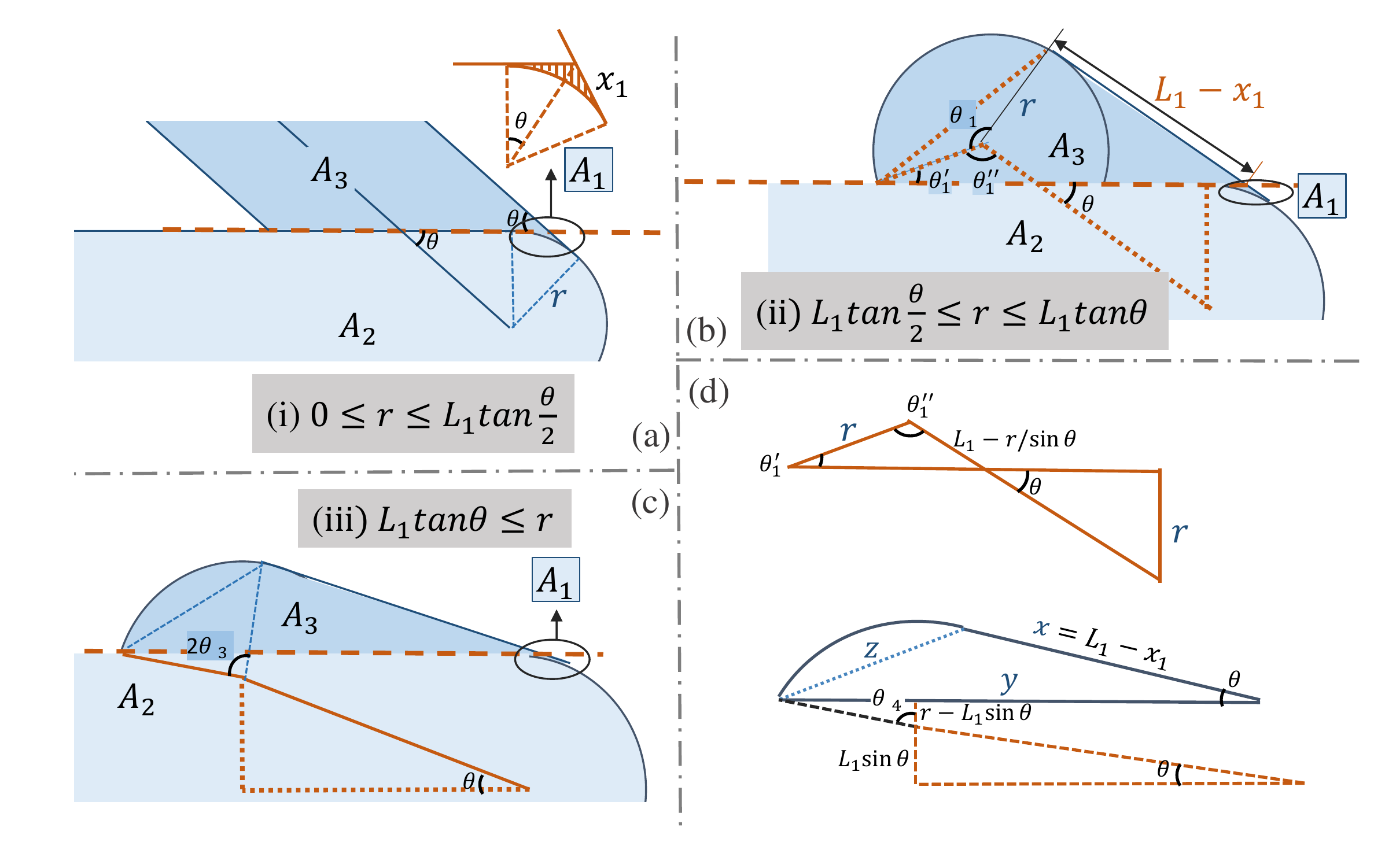}
		\caption{Poof of counting the area for $R_b$.}
		\label{App_proofRb}
	\end{figure}

To do so, we divide $A$ into 3 subarea, as shown in Fig.~\ref{App_proofRb}. Observing that in the case of different relationship between $\theta$ and $L_1$, the area of $A_1$ and the equation of computing  $A_2$ keep the same, while $A_3$ changes. Hence we mainly focus on $A_3$.
As Fig.~\ref{App_proofRb} shows,$A_2$ is consisted of a rectangle and two semicircles, thus the area is:
\begin{align}
	|A_2| = L_2\times 2 R_b+2\times\frac{1}{2}\pi R_b^2,
\end{align}
while $|A_1|$ is computing by the triangle subtract the chord, and $x_1$ is simply given by the triangular relationship,
\begin{align}
|A_1| &= 2R_b\times x_1-\frac{2\theta}{2\pi}\pi R_b^2,\nonumber\\
x_1 &= R_b\tan(\theta).
\end{align}
$|A_3|$ is slightly more complex compared with another two areas. We consider 3 possible cases as shown in Fig.~\ref{App_proofRb}(\textbf{a,b,c}) and in (\textbf{d}) we plot two important geometry relationships of (\textbf{b}) and (\textbf{c}).
The proof completes by using counting measure of Poisson distribution,
\begin{align}
	\mathbb{P}(R_b>r) = \mathbb{P}(\mathcal{N}(A) = 0) = \exp(-\lambda_b|A|).
\end{align}
\bibliographystyle{IEEEtran}
\bibliography{Ref6}
\end{document}